\documentclass[12pt,journal,onecolumn]{IEEEtran}

\usepackage{amssymb,amsmath,amsfonts,bm,epsfig,graphicx,theorem,latexsym}
\usepackage{rotating,setspace,latexsym,epsf,color,subfigure}
\usepackage{cite,times,authblk,epstopdf}
\usepackage{multicol}%
\usepackage{verbatim}%
\usepackage{enumerate}%------------------------
\usepackage{eucal}
\usepackage{caption}
\usepackage{stmaryrd,bbm,subfigure,units}

\newtheorem{theorem}{Theorem}

\newtheorem{corollary}{Corollary}

\newtheorem{lemma}{Lemma}

\newtheorem{remark}{Remark}

\newenvironment{Proof}[1]{\medskip\par\noindent{\bf Proof:\,}\,#1}{{\mbox{\,$\blacksquare$}\medskip\par}}

\makeatletter

\allowdisplaybreaks[1]

\newcommand{\bX}{\bold{X}}

\newcommand{\bx}{\bold{x}}

\newcommand{\bY}{\bold{Y}}
\newcommand{\by}{\bold{y}}
\newcommand{\bZ}{\bold{Z}}

\newcommand{\bz}{\bold{z}}
\newcommand{\bV}{\bold{V}}

\newcommand{\bU}{\bold{U}}
\newcommand{\bUh}{\hat{\bold{U}}}
\newcommand{\bu}{\bold{u}}
\newcommand{\buh}{\hat{\bold{u}}}
\newcommand{\tW}{\tilde{W}}
\newcommand{\tF}{\tilde{F}}
\newcommand{\tP}{\tilde{P}}

\newcommand{\setA}{\mathcal{A}}
\newcommand{\setB}{\mathcal{B}}
\newcommand{\setC}{\mathcal{C}}
\newcommand{\setD}{\mathcal{D}}

\newcommand{\setG}{\mathcal{G}}

\newcommand{\setW}{\mathcal{W}}

\newcommand{\setR}{\mathcal{R}}
\newcommand{\setS}{\mathcal{S}}

\newcommand{\setV}{\mathcal{V}}
\newcommand{\setX}{\mathcal{X}}
\newcommand{\setY}{\mathcal{Y}}
\newcommand{\setZ}{\mathcal{Z}}

\newcommand{\E}{{\mathbb{E}}}
\newcommand{\V}{{\mathbb{V}}}
\newcommand{\D}{{\mathbb{D}}}
\newcommand{\Prob}{{\mathbb{P}}}

\newcommand{\limitn}{{\underset{n\rightarrow\infty}\lim}}

\title{Generalizing Multiple Access Wiretap and Wiretap II Channel Models: Achievable Rates and Cost of Strong Secrecy}

\author{Mohamed~Nafea,~\IEEEmembership{Student Member,~IEEE,}
        and Aylin~Yener,~\IEEEmembership{Fellow,~IEEE}\\
{\normalsize Wireless Communications and Networking Laboratory (WCAN)\\
Electrical Engineering Department\\
The Pennsylvania State University, University Park, PA 16802.}\\
\em mnafea@psu.edu \qquad yener@engr.psu.edu

\thanks{This work was supported in part by NSF Grant CNS 13-14719. This paper was presented in part at the 2016 IEEE International Symposium on Information Theory \cite{nafea2016multiple} and the 2016 IEEE Information Theory Workshop \cite{nafea2016new2}.}}

\begin{document}

\IEEEoverridecommandlockouts

\maketitle
\vspace{-1cm} 

% ------------------------------ ABSTRACT -----------------------------
\begin{abstract}
In this paper, new two-user multiple access wiretap channel models are studied. First, the multiple access wiretap channel II with a discrete memoryless main channel, under three different wiretapping scenarios, is introduced. The wiretapper, as in the classical wiretap channel II model, chooses a fixed-length subset of the channel uses on which she obtains noise-free observations of one of the codewords, a superposition of the two codewords, or each of the two codewords. These thus extend the recently examined wiretap channel II with a noisy main channel to a multiple access setting with a variety of attack models for the wiretapper. Next, a new multiple access wiretap channel model, which further generalizes the multiple access wiretap channel II under the third wiretapping scenario, i.e., that which features the strongest adversarial model, is proposed. In this model, the wiretapper, besides choosing a subset of the channel uses to noiselessly observe the transmitted codeword symbols of both users, observes the remainder of the two codewords through a discrete memoryless multiple access channel. Achievable strong secrecy rate regions for all the proposed models are derived. Achievability is established by solving dual {\it{multi-terminal}} secret key agreement problems in the source model, and converting the solution to the original channel models using probability distribution approximation arguments. The derived achievable rate regions quantify the secrecy cost due to the additional capabilities of the wiretapper with respect to the previous multiple access wiretap models. 
\end{abstract}

\section{Introduction}\label{Int}
The wiretap channel II, in which the legitimate terminals communicate over a noiseless channel while the wiretapper has perfect access to a fixed fraction of her choosing of the transmitted bits, was introduced in \cite{WTCII_Wyner}. This model, while similar to a classical wiretap channel \cite{WTCWyner} with a noiseless main channel and a binary erasure channel to the wiretapper, models a more capable wiretapper who is able to select the positions of erasures. Using random partitioning and combinatorial arguments, \cite{WTCII_Wyner} has shown that the secrecy capacity of the wiretap channel II model does not decrease if the wiretapper is a passive observer with a binary erasure channel whose erasures are randomly chosen by nature, demonstrating the immunity of wiretap coding against a more capable adversary who is able to choose the erasure positions. 

Considerable amount of research on practical code design for secrecy has been motivated by the coset coding scheme devised in \cite{WTCII_Wyner}, see for example 
\cite{wei1991generalized,thangaraj2007applications,liu2007secure,aggarwal2009wiretap,bloch2015error}. However, for several decades, there has been no effort for generalizing the wiretap II model outside the special scenario of the noiseless main channel. Recently, \cite{nafea2015wiretap} has introduced a discrete memoryless  main channel to the wiretap channel II model, and derived inner and outer bounds for its capacity-equivocation region. Reference \cite{goldfeld2015semantic} has characterized the secrecy capacity of this model, showing that, once again, the secrecy capacity of the model does not decrease when the more capable wiretapper is replaced with an erasure channel.

More recently, \cite{nafea2017new} has introduced the generalized wiretap channel and identified its secrecy capacity. In this model, the main channel is a discrete memoryless channel while the wiretapper, besides noiselessly observing a subset of the transmitted codeword symbols of her choice, observes the remainder through a discrete memoryless channel. This new model subsumes both the classical wiretap channel \cite{CK} and the wiretap channel II with a discrete memoryless main channel \cite{nafea2015wiretap} as its special cases. The secrecy capacity of this generalized model quantifies the secrecy penalty of the additional capability at the wiretapper with respect to the previous wiretap models. Investigating the multi-terminal extensions of this new wiretap model is the natural next step, much like what happened with Wyner's wiretap channel, see for example \cite{yener2015wireless,tekin2008gaussian,tekin2008general,liang2008multiple,liu2008discrete,he2013role,he2011gaussian}, 

In this paper, we thus extend this new wiretap channel model to the multiple access scenario \cite{tekin2008general}. In particular, we first consider the special case of the multiple access wiretap channel II with a discrete memoryless main channel, and propose three different attack models for the wiretapper. In each of these models, the wiretapper chooses a fixed-length subset of the channel uses and observes erasures outside this subset. In the first wiretapping model, the wiretapper, in each position of the subset, decides to observe either the first or the second user's symbol. In the second model, the wiretapper observes a noiseless superposition of the two transmitted symbols in the positions of the subset, while in the third model, the wiretapper observes the transmitted symbols of both users.

The first attack model is a setting in which the wiretapper is able to tap one of the two transmissions but not both. For instance, if two transmitters are distant from each other, the wiretapper may need get close to one in order to obtain noise-free observations, and thus is able to tap one at a time. The second attack model mimics a medium that superposes both transmissions (e.g., wireless), where the attacker is close enough to both transmitters. In the third attack model, the wiretapper is able to tap both codewords individually, which can be interpreted as the wiretapper being able to obtain noiseless (partial) side information about both transmitted codewords.    

For each of these models, we derive an achievable strong secrecy rate region. Even though the third attack model, in which the wiretapper sees the transmitted symbols of {\it{both}} users, is stronger than the first, the ability of the wiretapper in the first model to choose which user's symbol to tap into results in identical achievable strong secrecy rate regions for the two models. That is, each transmitter designs their encoding according to the worst case scenario in which the wiretapper chooses to see his symbols in all positions of the subset. The achievable secrecy rate region for the second attack model is shown to be larger than the achievable secrecy rate region for the other two models, demonstrating the intrinsic cooperation introduced by superposition. 

After obtaining these insights, we generalize these models by replacing the wiretapper's erasures with noisy channel outputs as was done in \cite{nafea2017new} for the single user channel. In particular, we generalize the multiple access wiretap channel II with a discrete memoryless main channel, under the third wiretapping scenario, i.e., the strongest attack model, to the case when the wiretapper observes the remainder of the codewords of both users separately through a discrete memoryless channel. This model also generalizes the multiple access wiretap channel in \cite{tekin2008gaussian} to the case when the wiretapper is provided with a subset of noiseless observations of her choice for the transmitted symbols of both users. An achievable strong secrecy rate region, which quantifies the secrecy cost of the additional capability of the wiretapper in this model with respect to the multiple access wiretap channel in \cite{tekin2008gaussian,yassaee2010multiple}, is derived. 

Achievability of the strong secrecy rate regions for all the proposed models is established by muti-terminal extensions of methods in \cite{nafea2017new,ahlswede1993common,yassaee2014achievability}. In particular, for each of the proposed models, a corresponding dual multi-terminal secret key agreement problem in the source model is introduced. In this dual model, two independent sources wish to agree on two indepedent keys with a common decoder in the presence of a {\it{compound}} wiretapping source. We solve the problem in the dual source model, and convert the solution to the original channel model by means of deriving the joint distributions of the two problems to become almost identical, in the total variation distance sense. The technical challenge in the present paper lies in generalizing the tool utilized for establishing secrecy of the key in the dual source model from the single source case, \cite[Lemma 2]{nafea2017new}, to the case of two {\it{independent}} sources. This is done by adapting the lemma in order to establish all the {\it{corner}} (extreme) points of the rate region for the two keys, generated at the independent sources, such that the convergence rate for the probability of the two keys being independent from the wiretapper's observation is {\it{doubly-exponential}}. Time sharing between the resulting corner points produces the desired rate region. This doubly-exponential convergence rate is needed in order to {\it{exhaust}} the exponentially many possible strategies for the wiretapper \cite{goldfeld2015semantic,nafea2017new}. 

The remainder of the paper is organized as follows. Section \ref{ChannelModel} describes the channel models considered in this paper. Section \ref{MainResult} presents the main results. The proofs of the results are presented in Sections \ref{Achievability_Thm1} and \ref{Achievability_Thm2_3}. Section \ref{Con} concludes the paper.

\begin{figure}
\centering
\includegraphics[scale=0.8]{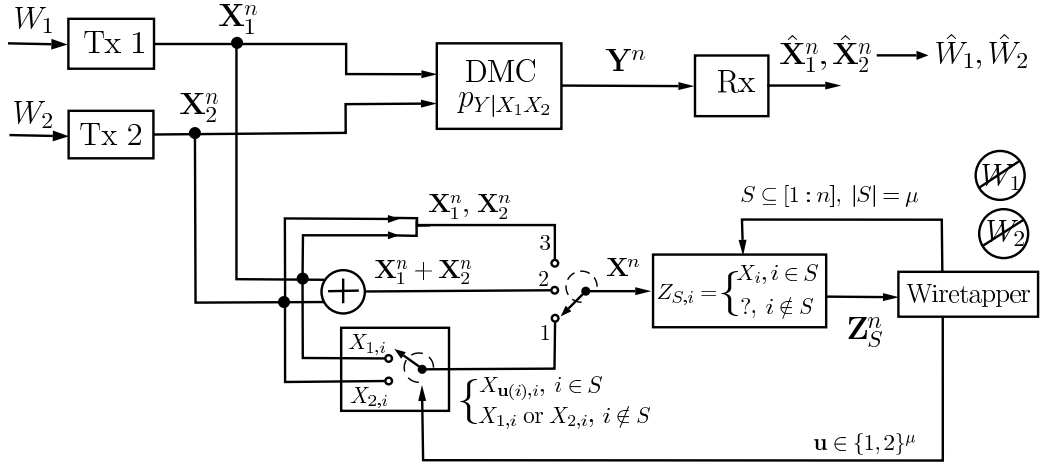}
\caption{The two-user multiple access wiretap channel II with a noisy main channel.}
\label{fig:sysmodel_1}
\end{figure}

\section{Channel Models}{\label{ChannelModel}}
We first remark the notation we use throughout the paper. Vectors are denoted by bold lower-case superscripted letters while their components are denoted by lower-case subscripted letters. A similar convention but with upper-case letters is used for random vectors and their components. Vector superscripts are omitted when the dimensions are clear from the context. $\setA_1\times\setA_2$ denotes the Cartesian product of the sets $\setA_1,\setA_2$. For random variables (vectors) and their components, we use $A_{[i:j]}$ to denote $\{A_i,\cdots,A_j\}$, where $i,j\in \mathbb{N}$, $i<j$. We also use $A_{S}\triangleq\{A_i\}_{i\in S}$ for $S\subseteq \mathbb{N}$. We use $\mathbbm{1}\{\setA\}$ to denote the indicator function of the event $\setA$. For $a,b\in \mathbb{R}$, $[a:b]$ denotes the set of integers $\{i\in\mathbb{N}: a\leq i\leq b\}$. We use upper-case letters to denote random\footnote{Random probability distribution $P_X$ describes a mapping from the random experiment to the simplex of probability distributions over $X$.} probability distributions, e.g., $P_X$. We use $p_X^U$ to denote a uniform distribution over the random variable $X$. The argument of the probability distribution is omitted when it is clear from its subscript. $\V(p_X,q_X)$ and $\D(p_X||q_X)$ denote the total variation distance and the Kullback-Leibler (K-L) divergence between the two probability distributions $p_X$ and $q_X$.  

Next, we describe the channel models we consider in this paper. In Section \ref{MAC_WTC_II}, we present the multiple access wiretap channel II with a noisy main channel under the three aforementioned attack models for the wiretapper. Section \ref{New_MAC_WTC} describes a new multiple access wiretap channel model that generalizes the strongest attack model in Section \ref{MAC_WTC_II}. 

\subsection{The Multiple Access Wiretap Channel II with a Noisy Main Channel}\label{MAC_WTC_II}
Consider the channel model in Fig. \ref{fig:sysmodel_1}. The main channel $\{\setX_1,\setX_2,\setY,p_{Y|X_1 X_2}\}$ is a discrete memoryless channel consisting of two finite input alphabets $\setX_1$ and $\setX_2$, a finite output alphabet $\setY$, and a transition probability distribution $p_{Y|X_1 X_2}$. Each transmitter wishes to reliably communicate an independent message to a common receiver and to keep it secret from the wiretapper. To do so, transmitter $j$ maps its message, $W_j$, uniformly distributed over $[1:2^{nR_{j}}]$, into the transmitted codeword $\bX_j^n=[X_{j,1},X_{j,2},\cdots,X_{j,n}]\in\setX_j^n$ using a stochastic encoder, $j=1,2$. The receiver observes the sequence $\bY^n=[Y_1,Y_2,\cdots,Y_n]\in\setY^n$ and outputs the estimates $\hat{W}_j,j=1,2,$ of the transmitted messages. As shown in Fig. \ref{fig:sysmodel_1}, we consider the following three models for the wiretapper channel.

\subsubsection{Model 1}\label{WT_Model_1}
This model is described in Fig. \ref{fig:sysmodel_1}, when the switch is on position $1$. The wiretapper chooses the subset $S_p\in\setS_p$ and the sequence $\bold{u}=[u_1,u_2,\cdots,u_{\mu}]\in\{1,2\}^\mu$, where 
$\setS_p\triangleq \{S_p\subseteq [1:n]:\;|S_p|=\mu\leq n \}$.
That is, $S_p$ represents the set of positions noiselessly tapped by the wiretapper and $\bold{u}$ represents her sequence of decisions to observe {\it{either the first or the second user}} codeword symbols. We define the fraction of the tapped symbols by the wiretapper as 
\begin{align}
\label{eq:alpha}
\alpha=\frac{\mu}{n},\qquad 0\leq \alpha\leq 1.
\end{align}
Let $S_p(k)$ and $\bold{u}(k)$ denote the $k$th elements of the subset $S_p$ and the sequence $\bold{u}$, where $k=1,2,\cdots,\mu$. Let $\setS$ be the set of all possible strategies for the wiretapper, where $\setS$ is defined as 
\begin{align}
\label{eq:setS_Model_A}
\setS\triangleq \left\{(S_p(k),\bold{u}(k)): S_p\in\setS_p,\;\bold{u}\in\{1,2\}^{\mu},\; k=1,2,\cdots,\mu\right\}.
\end{align}

For $S\in\setS$, the wiretapper observes $\bZ_{S}^n=[Z_{S,1}, Z_{S,2},\cdots,Z_{S,n}]\in {\mathcal{Z}^n}$, where
\begin{align}
\label{eq:Z_S_n_A}
Z_{S,i}=\begin{cases}
X_{j,i},\quad (i,j)\in S\\
?,\quad (i,j)\notin S,
\end{cases}
\end{align} 
and the alphabet $\setZ\triangleq\{\setX_1\cup\setX_2\}\cup\{?\}$.

\subsubsection{Model 2}\label{WT_Model_2}
The model is described in Fig. \ref{fig:sysmodel_1}, when the switch is on position $2$. The wiretapper chooses the subset $S\in\setS$, where we redfine the set $\setS$ as
\begin{align}
\label{eq:setS_Model_B}
\setS\triangleq\left\{S\subseteq [1:n]:\;|S|=\mu\leq n\right\}. 
\end{align}
The wiretapper then observes $\bZ_S^n=[Z_{S,1},Z_{S,2},\cdots,Z_{S,n}]\in {\mathcal{Z}^n}$, where
\begin{align}
\label{eq:Z_S_n_B}
Z_{S,i}=\begin{cases}
X_{1,i}+X_{2,i},\quad i\in S\\
?,,\quad i\notin S,
\end{cases}
\end{align} 
and $\setZ\triangleq\{\setX_1+\setX_2\}\cup\{?\}$. That is, the wiretapper observes noiseless {\it{superposition of the two users codeword symbols}} in the positions of the subset $S$, and erasures otherwise. The ratio $\alpha$ is defined as in (\ref{eq:alpha}). Note that in the definition of the set $\setZ$, we consider natural addition over the alphabets $\setX_1$ and $\setX_2$, i.e., $\setX_1+\setX_2\triangleq \{x_1+x_2:\; x_1\in\setX_1,\;x_2\in\setX_2\}$. 

\subsubsection{Model 3}\label{WT_Model_3}
The model is described in Fig. \ref{fig:sysmodel_1}, when the switch is on position $3$. The wiretapper chooses the subset $S\in\setS$, with $\setS$ defined as in (\ref{eq:setS_Model_B}), and observes $\bZ_S^n=[Z_{S,1},Z_{S,2},\cdots,Z_{S,n}]\in {\mathcal{Z}^n}$, where
\begin{align}
\label{eq:Z_S_n_C}
Z_{S,i}=\begin{cases}
\{X_{1,i},X_{2,i}\},\quad i\in S\\
?,,\quad i\notin S,
\end{cases}
\end{align} 
and $\setZ\triangleq \{\setX_1\times\setX_2\}\cup\{?\}$. That is, the wiretapper observes {\it{the transmitted codeword symbols of both users}} in the positions of the subset $S$, and erasures otherwise.

Next, we present a generalized multiple access wiretap channel model which extends the strongest attack model in Section \ref{WT_Model_3} to the case when the wiretapper sees noisy observations, instead of erasures, outside the subset she chooses.

\subsection{The Generalized Multiple Access Wiretap Channel}\label{New_MAC_WTC}
\begin{figure}
\centering
\includegraphics[scale=0.7]{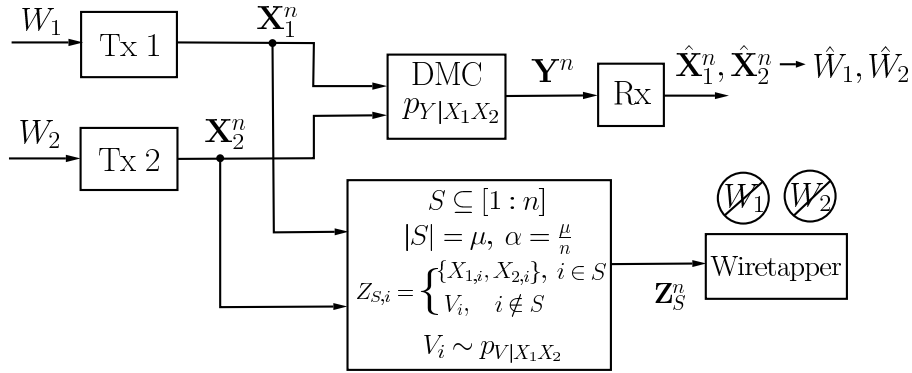}
\caption{The generalized two-user multiple access wiretap channel.}
\label{fig:sysmodel_2}
\end{figure}

Consider the channel model in Fig. \ref{fig:sysmodel_2}. The main channel in this model is identical to the main channel in Section \ref{MAC_WTC_II}. The wiretapper however chooses the subset $S\in\setS$, with $\setS$ defined as in (\ref{eq:setS_Model_B}), and observes $\bZ_S^n=[Z_{S,1},Z_{S,2},\cdots,Z_{S,n}]\in {\mathcal{Z}^n}$, where
\begin{align}
\label{eq:Z_S_n}
Z_{S,i}=\begin{cases}
\{X_{1,i},X_{2,i}\},\quad i\in S\\
V_i,\quad i\notin S.
\end{cases}
\end{align} 
$\bV^n=[V_1,V_2,\cdots,V_n]\in\setV^n$ is the $n$-letter output of the discrete memoryless multiple access channel $p_{V|X_1 X_2}$, $\setV$ is a finite alphabet, and $\setZ\triangleq\{\setX_1\times\setX_2\}\cup \setV$. 

For the channel models described in Sections \ref{MAC_WTC_II} and \ref{New_MAC_WTC}, an $(n,2^{nR_1},2^{nR_2})$ channel code $\setC_{n}\triangleq\{\setC_{1,n},\setC_{2,n}\}$ consists of two message sets $\setW_1=[1:2^{nR_1}]$, $\setW_2=[1:2^{nR_2}]$; two stochastic encoders $P_{\bX_1^n|W_1,\setC_{1,n}}$, $P_{\bX_2^n|W_2,\setC_{2,n}}$, and a decoder at the receiver. $(R_1,R_2)$ is an achievable strong secrecy rate pair if there exists a sequence of $(n,2^{nR_1},2^{nR_2})$ codes, $\{\mathcal{C}_n\}_{n\geq 1}$, such that 
\begin{align}
&\limitn\Prob\left(\bigcup_{j=1,2}(\hat{W}_j\neq W_j)\big|\mathcal{C}_n\right)=0,\\
&\text{and }\;\limitn\max_{S\in\setS} I(W_1,W_2;\bZ_S^n|\mathcal{C}_n)=0. 
\end{align}
Strong secrecy capacity region for the channel is the supremum of all achievable strong secrecy rate pairs $(R_1,R_2)$. In the following section, we describe the main results of this paper.

\section{Main Results}\label{MainResult}
We first present achievable strong secrecy rate regions for the two-user multiple access wiretap channel II with a discrete memoryless main channel, under the attack models for the wiretapper described in Sections \ref{WT_Model_1} and \ref{WT_Model_2}.

\begin{theorem}\label{thm:Thm1}
{\emph{For $0\leq \alpha\leq 1$, an achievable strong secrecy rate region for the multiple access wiretap channel II in Fig. \ref{fig:sysmodel_1} under the wiretapper model $1$, $\setR^{\rm{(1)}}(\alpha)$, is given by the convex hull of all rate pairs $(R_1,R_2)$ satisfying 
\begin{align}
\label{eq:Thm1_1} 
R_1&\leq I(U_1;Y|U_2)-\alpha I(U_1;X_1),\\
\label{eq:Thm1_2} 
R_2&\leq I(U_2;Y|U_1)-\alpha I(U_2;X_2),\\
\label{eq:Thm1_3}
R_1+R_2&\leq I(U_1,U_2;Y)-\alpha I(U_1,U_2;X_1,X_2),
\end{align}
for some distribution $p_{U_1X_1}p_{U_2X_2}$ which satisfies the Markov chains $U_1-X_1-Y$ and $U_2-X_2-Y$.}}
\end{theorem}

\begin{remark}
{\emph{The achievable strong secrecy rate region for the wiretapper model $1$ in Theorem \ref{thm:Thm1} is identical to the achievable region for the more capable wiretapper in model $3$, see Corollary \ref{cor:Cor_1}. When the wiretapper has the ability of choosing to observe either symbol in every tapped position, each user ought to design their transmission according to the worst case scenario in which the wiretapper decides to observe only his symbols in all the positions she taps. This results in an achievable rate region for the wiretapper model $1$ as when the wiretapper observes both users symbols in each position she taps.}}  
\end{remark}

\begin{theorem}\label{thm:Thm2}
{\emph{For $0\leq \alpha\leq 1$, an achievable strong secrecy rate region for the multiple access wiretap channel II in Fig. \ref{fig:sysmodel_1} under the wiretapper model $2$, $\setR^{\rm{(2)}}(\alpha)$, is given by the convex hull of all rate pairs $(R_1,R_2)$ satisfying  
\begin{align}
\label{eq:Thm2_1} 
R_1&\leq I(U_1;Y|U_2)-\alpha I(U_1;X_1+X_2),\\
\label{eq:Thm2_2} 
R_2&\leq I(U_2;Y|U_1)-\alpha I(U_2;X_1+X_2),\\
\label{eq:Thm2_3} 
R_1+R_2&\leq I(U_1,U_2;Y)-\alpha I(U_1,U_2;X_1+X_2),
\end{align}
for some distribution $p_{U_1X_1}p_{U_2X_2}$ which satisfies the Markov chains $U_1-X_1-Y$ and $U_2-X_2-Y$.}}
\end{theorem}

\begin{remark}
{\emph{The achievable strong secrecy rate region for the wiretapper models $1$ and $3$ is included in the achievable region for the wiretapper model $2$, i.e., $\setR^{\rm{(1)}}(\alpha)\subseteq \setR^{\rm{(2)}}(\alpha)$. This follows due to the Markov chains $U_1-X_1-(X_1+X_2)$; $U_2-X_2-(X_1+X_2)$, and $(U_1,U_2)-(X_1,X_2)-(X_1+X_2)$. By data processing inequality, we have 
\begin{align}
&I(U_j;X_j)\geq I(U_j;X_1+X_2),\qquad j=1,2,\\
&I(U_1,U_2;X_1,X_2)\geq I(U_1,U_2;X_1+X_2).
\end{align}}}
\end{remark}

Next, we present achievable strong secrecy rate regions for the generalized multiple access wiretap channel in Fig. \ref{fig:sysmodel_2}.

\begin{theorem}\label{thm:Thm3}
{\emph{For $0\leq \alpha\leq 1$, an achievable strong secrecy rate region for the generalized multiple access wiretap channel in Fig. \ref{fig:sysmodel_2}, $\setR(\alpha)$, is given by the convex hull of all rate pairs $(R_1,R_2)$ satisfying
\begin{align}
\label{eq:Thm3_1} 
R_1&\leq I(U_1;Y|U_2)-I(U_1;V)-\alpha I(U_1;X_1|V),\\
\label{eq:Thm3_2} 
R_2&\leq I(U_2;Y|U_1)-I(U_2;V)-\alpha I(U_2;X_2|V),\\
\label{eq:Thm3_3} 
R_1+R_2&\leq I(U_1,U_2;Y)-I(U_1,U_2;V)-\alpha I(U_1,U_2;X_1,X_2|V),
\end{align}
for some distribution $p_{U_1X_1}p_{U_2X_2}$ which satisfies the Markov chains $U_1-X_1-(Y,V)$ and $U_2-X_2-(Y,V)$.}}
\end{theorem}

\begin{corollary}\label{cor:Cor_1}
{\emph{For $0\leq \alpha\leq 1$, an achievable strong secrecy rate region for the multiple access wiretap channel II in Section \ref{WT_Model_3}, i.e., in Fig. \ref{fig:sysmodel_1} under the wiretapper model $3$, $\setR^{\rm{(3)}}(\alpha)$, is given by the convex hull of all rate pairs $(R_1,R_2)$ satisfying 
\begin{align}
\label{eq:Cor1_1} 
R_1&\leq I(U_1;Y|U_2)-\alpha I(U_1;X_1),\\
\label{eq:Cor1_2} 
R_2&\leq I(U_2;Y|U_1)-\alpha I(U_2;X_2),\\
\label{eq:Cor1_3} 
R_1+R_2&\leq I(U_1,U_2;Y)-\alpha I(U_1,U_2;X_1,X_2),
\end{align}
for some distribution $p_{U_1X_1}p_{U_2X_2}$ which satisfies the Markov chains $U_1-X_1-Y$ and $U_2-X_2-Y$.}}
\end{corollary}

Corollary \ref{cor:Cor_1} follows directly from Theorem \ref{thm:Thm3} by setting $V=\text{const.}$, i.e., the channel $p_{V|X_1 X_2}$ is an erasure channel with erasure probability one. The proofs for Theorems \ref{thm:Thm1}, \ref{thm:Thm2}, and \ref{thm:Thm3}, are provided in Sections \ref{Achievability_Thm1} and \ref{Achievability_Thm2_3}. 

\begin{remark}
{\emph{By setting the size of the subset $S$ to zero, i.e., $\alpha=0$, in Theorem \ref{thm:Thm3}, we obtain the achievable strong secrecy rate region in \cite[Theorem 1]{yassaee2010multiple} for the two user multiple access wiretap channel. The same region was derived under a weak secrecy criterion in \cite{tekin2008general,simeone2009cognitive}.}}
\end{remark}

\section{Proof for Theorem \ref{thm:Thm1}}{\label{Achievability_Thm1}}
The achievability proof for Theorem \ref{thm:Thm1} follows the same key steps as in \cite{nafea2017new}, with the need of extending the technique to address the multi-terminal setting as will be explained shortly. In particular, we first assume the availability of common randomness at all terminals of the original channel model. We then define a dual {\it{multi-terminal}} secret key agreement problem in the source model, which introduces a set of random variables similar to those introduced by the original problem with the assumed common randomness. We then solve for rate conditions which result in the induced joint distributions from the two models to be almost identical in the total variation distance sense. We also provide rate conditions which satisfy certain reliability and secrecy (independence) conditions in the source model. Next, we use the closeness of the induced joint distributions to show that, under the same rate conditions, the desired reliability and secrecy properties in the original channel model are satisfied. Finally, we eliminate the common randomness from the channel model by conditioning on a certain instance of that randomness.

The outline of achievability is hence threefold: (i) Reliability of the keys in the dual source model, (ii) Security of the keys in the dual source model, and (iii) Closeness of the induced joint distributions. Reliability of the keys follows from Slepian-Wolf source coding theorem for multiple sources \cite[Theorem 10.3]{el2011network}. Closeness of joint distributions, and converting the reliability and security conditions from the dual model to the original problem, are ensured by deriving an {\it{exponential}} convergence rate for the average total variation distance between the two distributions. This is done using a rather straightforward generalization of \cite[Lemma 1]{nafea2017new}. 

The main challenge in the proof lies in ensuring security for the keys in the dual source model, which requires {\it{doubly-exponential}} convergence rate for the probability of the {\it{two keys}} being uniform and independent from the wiretapper's observation, in the Kullback-Leibler divergence sense. The double-exponential convergence is needed in order to ensure security against the exponentially many possible strategies for the wiretapper. This is established by adapting the lemma derived for the single source case in \cite{nafea2017new} so that we derive the corner points of the rate region, for the two keys, that satisfies the doubly-exponential convergence. Time sharing between these corner points hence results in the desired rate region. 

Let us first fix the distribution $p_{U_1 X_1}p_{U_2 X_2}=p_{U_1}p_{U_2}p_{X_1|U_1} p_{X_2|U_2}$. Let $p_{Y|U_1 U_2}$ be the distribution resulting from concatenating the discrete memoryless channels $p_{Y|X_1 X_2}$ and $p_{X_1 X_2|U_1 U_2}=p_{X_1|U_1}p_{X_2|U_2}$, where $p_{Y|X_1 X_2}$ is the main channel transition probability distribution for the model in Section \ref{MAC_WTC_II}. That is, 
\begin{align}
p_{Y|U_1 U_2}(y|u_1,u_2)=\sum_{x_1,x_2\in\setX_1\times\setX_2} p_{X_1|U_1}(x_1|u_1)\;p_{X_2|U_2}(x_2|u_2)\;p_{Y|X_1X_2}(y|x_1,x_2).
\end{align}
We describe the following two protocols, each of which introduces a set of random variables and induces a joint distribution over them. We precisely identify the joint distribution induced by each protocol. 

\begin{figure}
    \centering
	\includegraphics[scale=1]{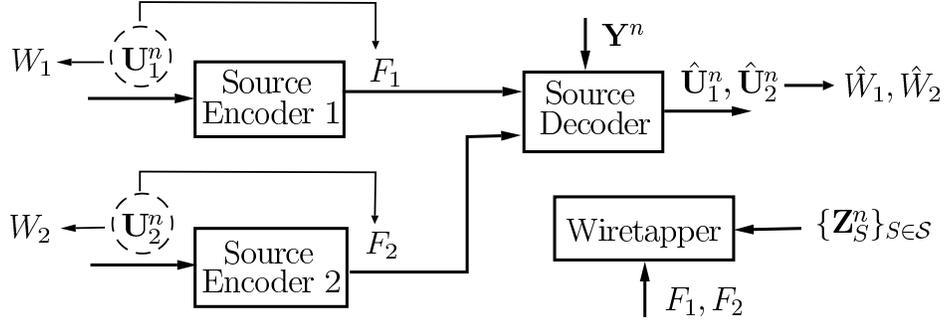}
	\caption{Protocol A: Multi-terminal secret key agreement problem in the source model.}
	\label{fig:Protocol A}
\end{figure}

{\it{Protocol A:}} This protocol describes a multi-terminal secret key agreement problem in the source model as shown in Fig. \ref{fig:Protocol A}. Let $\bU_1^n,\bU_2^n,\bY^n$ be independent and identically distributed (i.i.d.) sequences according to the distribution $p_{U_1}p_{U_2}p_{Y|U_1 U_2}$. Source encoder $j$ observes the sequence $\bU_j$, $j=1,2$. The sequence $\bU_j$ is randomly and independently binned into the two indices $W_j=\setB^{(j)}_1(\bX_j)$ and $F_j=\setB^{(j)}_2(\bX_j)$, where $\setB^{(j)}_1$ and $\setB^{(j)}_2$ are independent and uniformly distributed over $[1:2^{nR_j}]$ and $[1:2^{n\tilde{R}_j}]$, respectively. The bins $F_j,j=1,2,$ represent the public messages transmitted noiselessly to the common decoder and perfectly accessed by the wiretapper. The bins $W_j,j=1,2,$ represent the independent confidential keys generated at the two encoders. The decoder observes the i.i.d. sequence $\bY$ and the public messages $F_1,F_2$, and outputs the estimates $\bUh_1$, $\bUh_2$, $\hat{W}_1$, $\hat{W}_2$. 

Let $\setS$ and $\bZ_S$, for all $S\in\setS$, be defined as in (\ref{eq:setS_Model_A}) and (\ref{eq:Z_S_n_A}). The wiretapper chooses the strategy $S\in\setS$ whose realization is unknown to the legitimate terminals. The wiretapper can thus be represented by the source ${\bZ_S}\triangleq\{\mathcal{Z},{p_{\bZ_S}}, S\in\setS\}$ whose distribution is only known to belong to the finite class $\{p_{\bZ_S}\}_{S\in\setS}$; the cardinality of the set $\setS$ of all possible wiretapper's strategies for the attack model $1$ is upper bounded as 
\begin{align}
\label{eq:setS_Model_A_card}
|\setS|=\binom{n}{\mu}\times 2^{\mu}=\binom{n}{\alpha n}\times 2^{\alpha n}< 2^n\times 2^{\alpha n}= 2^{(1+\alpha) n}.
\end{align}  

Protocol A hence introduces the random variables $W_{[1:2]}, F_{[1:2]},\bU_{[1:2]},\bY,\bZ_S,\bUh_{[1:2]},\hat{W}_{[1:2]}$. The induced distribution over these variables is given by 
\begin{align}
\label{eq:P_tilde_0}
&\tP_{W_{[1:2]}F_{[1:2]}\bU_{[1:2]}\bY\bZ_S\bUh_{[1:2]}}=
p_{\bU_{[1:2]}\bY\bZ_S} \tP_{W_{[1:2]}F_{[1:2]}|\bU_{[1:2]}}\tP_{\bUh_{[1:2]}|\bY F_{[1:2]}}\\
\label{eq:P_tilde_1}
&=p_{\bU_{[1:2]}\bY\bZ_S}\tP_{\bUh_{[1:2]}|\bY F_{[1:2]}}\mathbbm{1}\left\{\setB_1^{(j)}(\bU_j)=W_j,\setB_2^{(j)}(\bU_j)=F_j,\forall j=1,2\right\}\\
\label{eq:P_tilde_2}
&=\tP_{W_{[1:2]} F_{[1:2]}}\tP_{\bU_{[1:2]}|W_{[1:2]} F_{[1:2]}}\;p_{\bY \bZ_S|\bU_{[1:2]}}\tP_{\bUh_{[1:2]}|\bY F_{[1:2]}}.
\end{align}
 
{\it{Protocol B:}} This protocol is described as the original channel model in Section \ref{WT_Model_1}, with assuming the availability of common randomness $F_1$, $F_2$, at all terminals. $F_1$ and $F_2$ are independent, uniformly distributed over $[1:2^{n\tilde{R}_1}]$ and $[1:2^{n\tilde{R}_2}]$, and independent from all other random variables. We utilize here the encoders and decoder  in (\ref{eq:P_tilde_2}). That is,
\begin{align}
P_{\bU_{[1:2]}|W_{[1:2]}F_{[1:2]}}=\tP_{\bU_{[1:2]}|W_{[1:2]}F_{[1:2]}},\quad \text{ and }\quad P_{\bUh_{[1:2]}|\bY F_{[1:2]}}=\tP_{\bUh_{[1:2]}|\bY F_{[1:2]}}.
\end{align}
The induced joint distribution for protocol B is thus given by 
\begin{align}
\label{eq:P_1}
P_{W_{[1:2]}F_{[1:2]}\bU_{[1:2]}\bY\bZ_S\bUh_{[1:2]}}=p_{W_{[1:2]}}^Up_{F_{[1:2]}}^U\tP_{\bU_{[1:2]}|W_{[1:2]}F_{[1:2]}}p_{\bY \bZ_S|\bU_{[1:2]}}
\tP_{\bUh_{[1:2]}|\bY F_{[1:2]}}.
\end{align}

\begin{remark}
{\emph{We have ignored the $\hat{W}$ variables from the joint distributions in (\ref{eq:P_tilde_2}) and (\ref{eq:P_1}) at this stage, as we will introduce them later as deterministic functions of the $\bUh$ random vectors, after fixing the binning functions.}}
\end{remark}

\begin{remark}
{\emph{Notice that $\tP_{\bU_{[1:2]}|W_{[1:2]}F_{[1:2]}}$ factorizes as $\tP_{\bU_1|W_1 F_1}\tP_{\bU_2|W_2 F_2}$. That is, the common randomness $F_i$ available at the $j$th transmitter, $i,j=1,2, i\neq j$, is not utilized to generate $\bU_j$. The common randomness $F_i,i=1,2,$ represents the realization of transmitter $i$'s codebook, which is known at all terminals. However, the transmitted codeword at one transmitter does not depend on the codebook of the other transmitter.}}
\end{remark}

\begin{remark}
{\emph{The induced joint distributions from the two protocols in (\ref{eq:P_tilde_2}) and (\ref{eq:P_1}) are random due to the random binning of $\bU_1^n$ and $\bU_2^n$.}} 
\end{remark}

Before continuing with the proof, we state the following lemmas.  

\subsection{Useful Lemmas}\label{Lemmas}
By comparing the joint distributions for protocols A and B in (\ref{eq:P_tilde_2}) and (\ref{eq:P_1}), we find that they only differ in the distribution for $W_{[1:2]}$ and $F_{[1:2]}$. In particular, $W_{[1:2]}$ and $F_{[1:2]}$ are independent and uniformly distributed in protocol B, while their distribution in protocol A is determined by the random binning of $\bU_{1}$ and $\bU_2$. The following lemma is a one-shot result which provides conditions on the binning rates such that the random binning of $\bU_1$ and $\bU_2$ described in protocol A results in a distribution for the bins that is close, in the total variation distance sense, to independent uniform distributions. The convergence rate provided by the lemma, which is {\it{exponential}}, is  needed for converting the secrecy (independence) condition, established for the source model in protocol A, to the original channel model in protocol B. 

\begin{lemma}\label{lemma1}
{\emph{Let $X_1\triangleq\{\setX_1,p_{X_1}\}$ and $X_2\triangleq\{\setX_2,p_{X_2}\}$ be two independent sources. The source $X_j, j=1,2,$ is randomly binned into the two indices $W_j=\setB^{(j)}_1(X_j)$ and $F_j=\setB^{(j)}_2(X_j)$, where $\setB^{(j)}_1$ and $\setB^{(j)}_2$ are independent and uniformly distributed over $[1:\tW_j]$ and $[1:\tF_j]$. Let $\setB\triangleq\left\{\setB^{(j)}_1(x_j),\setB^{(j)}_2(x_j): x_j\in\setX_j,j=1,2\right\}$, and for $\gamma_j>0, j=1,2,$ define 
\begin{align}
\label{eq:lemma1_1}
\setD_{\gamma_j}\triangleq \left\{x_j\in\setX_j: \log \frac{1}{p_{X_j}(x_j)}>\gamma_j\right\}.
\end{align}
Then, we have
\begin{align}
\label{eq:lemma1}
\E_{\setB}\left(\V\left(P_{W_{[1:2]}F_{[1:2]}},p_{W_{[1:2]}}^U p_{F_{[1:2]}}^U\right)\right)\leq 
\sum_{j=1}^2\left(\Prob_{P_{X_j}}\left(X_j\notin \setD_{\gamma_j}\right)+\frac{1}{2}\sqrt{\tW_j\tF_j2^{-\gamma_j}}\right), 
\end{align} 
where $P$ is the induced distribution over $W_{[1:2]}$ and $F_{[1:2]}$.}} 
\end{lemma}

\begin{Proof} 
Lemma \ref{lemma1} is a generalization of \cite[Lemma 1]{nafea2017new}. In particular, using the triangle inequality,
\begin{align}
\label{eq:lemma1_proof_1}
\V&\left(P_{W_{[1:2]}F_{[1:2]}},p_{W_{[1:2]}}^Up_{F_{[1:2]}}^U\right)=\V\left(P_{W_1 F_1}P_{W_2 F_2},p_{W_{[1:2]}}^Up_{F_{[1:2]}}^U\right)\\
\label{eq:lemma1_proof_2}
&\qquad\leq \V\left(P_{W_1 F_1}P_{W_2 F_2},p_{W_1}^Up_{F_1}^UP_{W_2F_2}\right)+\V\left(p_{W_1}^Up_{F_1}^UP_{W_2F_2},p_{W_{[1:2]}}^Up_{F_{[1:2]}}^U\right)\\
\label{eq:lemma1_proof_3}
&\qquad=\sum_{j=1,2}\V\left(P_{W_jF_j},p_{W_j}^Up_{F_j}^U\right),
\end{align}
where (\ref{eq:lemma1_proof_1}) follows since $X_1$ and $X_2$ are independent, and hence $\{W_1,F_1\}$ and $\{W_2,F_2\}$ are independent as well. Using \cite[Lemma 1]{nafea2017new}, we have, for $j=1,2,$
\begin{align}
\label{eq:lemma1_proof_4}
\E_{\setB}\left(\V\left(P_{W_j F_j},p_{W_j}^U p_{F_j}^U\right)\right)\leq 
\Prob_{P_{X_j}}\left(X_j\notin \setD_{\gamma_j}\right)+\frac{1}{2}\sqrt{\tW_j\tF_j2^{-\gamma_j}}, 
\end{align}
which completes the proof for Lemma \ref{lemma1}.
\end{Proof}

Lemma \ref{lemma2} below is again a one-shot result which provides rate conditions for a certain secrecy (independence) condition in the source model. In particular, the lemma provides a {\it{doubly-exponential}} convergence rate for the probability of the confidential keys $W_{[1:2]}$ and the public messages $F_{[1:2]}$ being independent, uniformly distributed, and all independent from the wiretapper's observation $\bZ_S$. This doubly-exponential convergence is utilized, along with the union bound, to guarantee secrecy against the exponentially many choices for the wiretapper. 
   
\begin{lemma}\label{lemma2} 
{\emph{Let $X_1\triangleq\{\setX_1,p_{X_1}\}$ and $X_2\triangleq\{\setX_2,p_{X_2}\}$ be two sources, both are correlated with the source $\{Z_{S}\}\triangleq\left\{\setZ,p_{Z_S}\right\}, S\in\setS$. The alphabets $\setX_1,\setX_2,\setZ$, and $\setS$, are finite. For $j=1,2,$ the source $X_j$ is randomly binned into the two indices $W_j$ and $F_j$ as in Lemma \ref{lemma1}. For $\gamma_j,\gamma_{ij}>0, i,j=1,2, i\neq j,$ and for any $S\in\setS$, define
\begin{align}
\label{eq:lemma2_1}
&\setD_{j}^S\triangleq \left\{(x_{[1:2]},z)\in\setX_1\times \setX_2\times\setZ:(x_j,z)\in\setD_{\gamma_j}^S,(x_{[1:2]},z)\in\setD_{\gamma_{ij}}^S\right\},\\
\label{eq:lemma2_2}
&\text{where}\quad\setD_{\gamma_j}^S\triangleq \left\{(x_j,z)\in\setX_j\times\setZ:\log\frac{1}{p_{X_j|Z_S}(x_j|z)}>\gamma_j\right\},\\
\label{eq:lemma2_3}
&\text{and}\quad \setD_{\gamma_{ij}}^S\triangleq\left\{(x_{[1:2]},z)\in\setX_1\times \setX_2\times\setZ: \log\frac{1}{p_{X_i|X_jZ_S}(x_i|x_j,z)}>\gamma_{ij}\right\}.
\end{align} 
If there exists a $\delta\in(0,\frac{1}{2})$ such that for $j=1,2,$ and for all $S\in\setS$, we have
\begin{align}
\label{eq:lemma2_4}
\Prob_{p_{X_{[1:2]}Z_S}}\left((X_{[1:2]},Z_S)\in\setD_j^S\right)\geq 1-\delta^2,
\end{align}
then, we have, for every $\epsilon\in[0,1]$, that
\begin{align}\label{eq:lemma2}
&\nonumber\Prob_{\setB}\left(\underset{S\in\setS}\max\; \D(P_{W_{[1:2]}F_{[1:2]}Z_S}||p_{W_{[1:2]}}^Up_{F_{[1:2]}}^U p_{Z_S})\geq 2\tilde{\epsilon}\right)\\
&\qquad\qquad\leq|\setS||\setZ|\underset{i,j=1,2,i\neq j}\min \left\{\exp\left(\left(\frac{-\epsilon^2 (1-\delta)2^{\gamma_j}}{3\tW_j\tF_j}\right)\right)+\exp\left(\left(\frac{-\epsilon^2 (1-\delta)2^{\gamma_{ij}}}{3\tW_i\tF_i}\right)\right)\right\},
\end{align}
where $P$ is the induced distribution over $W_{[1:2]}$ and $F_{[1:2]}$,
\begin{align}
\label{eq:lemma2_5}
\tilde{\epsilon}=\max_{j=1,2}\left\{\epsilon+(\delta+\delta^2)\log (\tW_j\tF_j)+H_b(\delta^2)\right\},
\end{align}
and $H_b$ is the binary entropy function.}}
\end{lemma}
\begin{Proof}
See the Appendix.
\end{Proof}

\begin{remark}
{\emph{In applying Lemmas \ref{lemma1} and \ref{lemma2} to the source model in protocol A, we utilize the version of Hoeffding's inequality in \cite[Theorem 2]{hoeffding1963probability}, \cite[Lemma 3]{nafea2017new}. In addition, after showing that the reliability and secrecy properties established for the source model hold as well for the channel model in protocol B, we utilize the selection lemma, \cite[Lemma 2.2]{bloch2011physical}, in order to prove the existence of a binning realization such that both properties are still satisfied for the channel model. It is also utilized to eliminate the common randomness $F_{[1:2]}$ from the channel model.}}
\end{remark}

\subsection{Proof}\label{Proof_Thm1}
We first apply Lemma \ref{lemma1} to the source model in protocol A to establish the closeness of the induced joint distributions from the two protocols. In Lemma \ref{lemma1}, set $X_{j}=\bU_{j}$, $\tW_j=2^{nR_j}$, and $\tF_j=2^{n\tilde{R}_j}$, for $j=1,2$; $\bU_j,\tW_j,\tF_j$ are defined as in protocol A. Let $\setD_{\gamma_j}$ be defined as in (\ref{eq:lemma1_1}) with $X_j=\bU_j$ for $j=1,2$. For $\epsilon_j>0, j=1,2,$ choose $\gamma_j=n(1-\epsilon_j)H(U_j)$.  Without loss of generality, assume that for all $\bu_j, j=1,2,$ $p_{\bU_j}(\bu_j)>0$. Using Hoeffding's inequality, we have 
\begin{align}
\label{eq:proof_Thm1_1}
&\Prob_{p_{\bU_j}}\left(\bU_j\notin\setD_{\gamma_j}\right)=\Prob\left(\log\frac{1}{p_{\bU_j}(\bU_j)}\leq \gamma_j\right)\\
\label{eq:proof_Thm1_2}
&=\Prob\left(\sum_{k=1}^n \log\frac{1}{p_{\bU_j}(U_{j,k})}\leq n(1-\epsilon_j)H(U_j)\right)\leq \exp(-\beta_j n),
\end{align} 
where $\beta_j>0$. By substituting the choices for $\tW_j,\tF_j,\gamma_j,$ and (\ref{eq:proof_Thm1_2}) in (\ref{eq:lemma1}), as long as 
\begin{align}
\label{eq:proof_Thm1_3}
&R_1+\tilde{R}_1<(1-\epsilon_1)H(U_1)\\
\label{eq:proof_Thm1_4}
&R_2+\tilde{R}_2<(1-\epsilon_2)H(U_2),
\end{align}
there exists a $\beta>0$ such that  
\begin{align}
\label{eq:proof_Thm1_5}
\E_{\setB}\left(\V\left(\tP_{W_{[1:2]}F_{[1:2]}},p_{W_{[1:2]}}^Up_{F_{[1:2]}}^U\right)\right)\leq 4\exp(-\beta n).
\end{align}

Using (\ref{eq:P_tilde_2}), (\ref{eq:P_1}), and (\ref{eq:proof_Thm1_5}), we have 
\begin{align}
\label{eq:proof_Thm1_5_1}
\nonumber\E_{\setB}&\left(\V\left(\tP_{W_{[1:2]}F_{[1:2]}\bU_{[1:2]}\bY\bZ_S\bUh_{[1:2]}},P_{W_{[1:2]}F_{[1:2]}\bU_{[1:2]}\bY\bZ_S\bUh_{[1:2]}}\right)\right)\\
&=\E_{\setB}\left(\V\left(\tP_{W_{[1:2]}F_{[1:2]}},p_{W_{[1:2]}}^Up_{F_{[1:2]}}^U\right)\right)\leq 4\exp(-\beta n).
\end{align}

Next, we establish a reliability condition for the source model in protocol A. We utilize a Slepian-Wolf decoder \cite{slepian1973noiseless}, which implies that \cite[Theorem 10.3]{el2011network}
\begin{align}
\label{eq:proof_Thm1_6}
\limitn\E_{\setB}\left(\Prob_{\tP}(\bUh_{[1:2]}\neq \bU_{[1:2]})\right)=0,
\end{align}
as long as
\begin{align}
\label{eq:proof_Thm1_8}
&\tilde{R}_1\geq H(U_1|U_2,Y),\\
\label{eq:proof_Thm1_9}
&\tilde{R}_2\geq H(U_2|U_1,Y),\\
\label{eq:proof_Thm1_10}
&\tilde{R}_1+\tilde{R}_2\geq H(U_1,U_2|Y).
\end{align}
Using (\ref{eq:proof_Thm1_6}) and \cite[Lemma 1]{yassaee2014achievability}, which is a variation on the Slepian-Wolf source coding theorem, we have, for all $S\in\setS$,
\begin{align}
\nonumber \limitn\E_{\setB}&\left(\V\left(\tP_{W_{[1:2]}F_{[1:2]}\bU_{[1:2]}\bY\bZ_S\bUh_{[1:2]}},\tP_{W_{[1:2]}F_{[1:2]}\bU_{[1:2]}\bY\bZ_S}\mathbbm{1}\{\bUh_{[1:2]}=\bU_{[1:2]}\}\right)\right)\\
\label{eq:proof_Thm1_11}
&=\limitn\E_{\setB}\left(\Prob_{\tP}(\bUh_{[1:2]}\neq \bU_{[1:2]})\right)=0.
\end{align}

Next, we use Lemma \ref{lemma2} to establish the secrecy condition for the source model in protocol A. In Lemma \ref{lemma2}, for $j=1,2,$ set $X_{j}=\bU_{j}$, $\tW_j=2^{nR_j}$, $\tF_j=2^{n\tilde{R}_j}$, $Z_S=\bZ_S$, for all $S\in\setS$, where $\bU_j,\setS,\bZ_S$ are defined as in protocol A. In addition, let $\setD_j^S,\setD_{\gamma_j}^S$, and $\setD_{\gamma_{ij}}^S$ be defined as in (\ref{eq:lemma2_1})-(\ref{eq:lemma2_3}), with $X_j=\bU_j$ and $Z_S=\bZ_S$.

For $S\in\setS$, define $\overline{S}_j\triangleq\{k:(k,j)\in S\}$. That is, $\overline{S}_j$ is the set of positions in which the wiretapper observes the $j$th transmitter's symbols. For $j=1,2,$ let $|\overline{S}_j|=\mu_j$, and hence $\mu_1+\mu_2=\mu$. Thus, we have 
\begin{align}
\label{eq:proof_Thm1_12}
H(\bU_1|\bZ_S)&=H(\bU_1|\bX_{1,\overline{S}_1},\bX_{2,\overline{S}_2})=H(\bU_{1,\overline{S}_1},\bU_{1,\overline{S}_1^c}|\bX_{1,\overline{S}_1},\bX_{2,\overline{S}_2})\\
\label{eq:proof_Thm1_13}
&=H(\bU_{1,\overline{S}_1}|\bX_{1,\overline{S}_1},\bX_{2,\overline{S}_2})+H(\bU_{1,\overline{S}_1^c}|\bX_{1,\overline{S}_1},\bX_{2,\overline{S}_2},\bU_{1,\overline{S}_1})\\
\label{eq:proof_Thm1_14}
&=H(\bU_{1,\overline{S}_1}|\bX_{1,\overline{S}_1})+H(\bU_{1,\overline{S}_1^c})=\mu_1 H(U_1|X_1)+(n-\mu_1) H(U_1)\\
\label{eq:proof_Thm1_15}
H(\bU_2|\bZ_S)&=\mu_2 H(U_2|X_2)+(n-\mu_2) H(U_2)\\
\label{eq:proof_Thm1_16}
H(\bU_1|\bU_2,&\bZ_S)=H(\bU_1|\bU_2,\bX_{1,\overline{S}_1},\bX_{2,\overline{S}_2})=H(\bU_{1,\overline{S}_1},\bU_{1,\overline{S}_1^c}|\bU_2,\bX_{1,\overline{S}_1},\bX_{2,\overline{S}_2})\\
\label{eq:proof_Thm1_17}
&=H(\bU_{1,\overline{S}_1}|\bU_2,\bX_{1,\overline{S}_1},\bX_{2,\overline{S}_2})+H(\bU_{1,\overline{S}_1^c}|\bU_2,\bX_{1,\overline{S}_1},\bX_{2,\overline{S}_2},\bU_{1,\overline{S}_1})\\
\label{eq:proof_Thm1_18}
&=H(\bU_{1,\overline{S}_1}|\bX_{1,\overline{S}_1})+H(\bU_{1,\overline{S}_1^c})=\mu_1 H(U_1|X_1)+(n-\mu_1)H(U_1)\\
\label{eq:proof_Thm1_19}
H(\bU_2|\bU_1,&\bZ_S)=\mu_2 H(U_2|X_2)+(n-\mu_2)H(U_2),
\end{align}
where (\ref{eq:proof_Thm1_14}) follows since $\{\bU_{1,\overline{S}_1},\bX_{1,\overline{S}_1}\}$ are independent from $\bX_{2,\overline{S}_2}$, and $\bU_{1,\overline{S}_1^c}$ is independent from $\{\bX_{1,\overline{S}_1},\bX_{2,\overline{S}_2},\bU_{1,\overline{S}_1}\}$, since $\bU_1$ is an i.i.d. sequence and $p_{X_1|U_1}$ is a discrete memoryless channel. Similarly, (\ref{eq:proof_Thm1_18}) follows since $\{\bU_{1,\overline{S}_1},\bX_{1,\overline{S}_1}\}$ are independent from $\{\bU_2,\bX_{2,\overline{S}_2}\}$, and $\bU_{1,\overline{S}_1^c}$ is independent from $\{\bU_2,\bX_{1,\overline{S}_1},\bX_{2,\overline{S}_2},\bU_{1,\overline{S}_1}\}$.

In addition, for the tuples $(\bx_{[1,2]},\bz)$ with $p_{\bX_j|\bZ_S}(\bx_j|\bz)>0$ and $p_{\bX_i|\bX_j\bZ_S}(\bx_i|\bx_j,\bz)>0$, where $i,j=1,2, i\neq j$, we have, for all $S\in\setS$, that
\begin{align}
\nonumber p_{\bU_j|\bZ_S}(\bu_j|\bz)&=p(\bu_{j,\overline{S}_j},\bu_{j,\overline{S}_j^c}|\bx_{j,\overline{S}_j},\bx_{i,\overline{S}_i})=p(\bu_{j,\overline{S}_j}|\bx_{j,\overline{S}_j},\bx_{i,\overline{S}_i})\;p(\bu_{j,\overline{S}_j^c}|\bu_{j,\overline{S}_j},\bx_{j,\overline{S}_j},\bx_{i,\overline{S}_i})\\
\label{eq:proof_Thm1_20}
&=p(\bu_{j,\overline{S}_j}|\bx_{j,\overline{S}_j})\;p(\bu_{j,\overline{S}_j^c})=\prod_{k\in \overline{S}_j}p(u_{j,k}|x_{j,k})\prod_{k\in \overline{S}_j^c}p(u_{j,k}),\\
\label{eq:proof_Thm1_21}
p_{\bU_i|\bU_j\bZ_S}(\bu_i|\bu_j,\bz)&=p(\bu_{i,\overline{S}_i}|\bx_{i,\overline{S}_i})\;p(\bu_{i,\overline{S}_i^c})=\prod_{k\in \overline{S}_i}p(u_{i,k}|x_{i,k})\prod_{k\in \overline{S}_i^c}p(u_{i,k}).
\end{align}

For $i,j=1,2, i\neq j,$ and $\tilde{\epsilon}_j>0$, let
\begin{align}
\label{eq:proof_Thm1_22}
&\gamma_{j}=(1-\tilde{\epsilon}_j)\;\underset{S\in\setS}\min \;H(\bU_j|\bZ_S)=(1-\tilde{\epsilon}_j)[\mu H(U_j|X_j)+(n-\mu)H(U_j)],\\
\label{eq:proof_Thm1_23}
&\gamma_{ij}=(1-\tilde{\epsilon}_j)\;\underset{S\in\setS}\min \;H(\bU_i|\bU_j,\bZ_S)=(1-\tilde{\epsilon}_j)[\mu H(U_i|X_i)+(n-\mu )H(U_i)], 
\end{align}
where (\ref{eq:proof_Thm1_22}) and (\ref{eq:proof_Thm1_23}) follow from (\ref{eq:proof_Thm1_14}), (\ref{eq:proof_Thm1_15}), (\ref{eq:proof_Thm1_18}), (\ref{eq:proof_Thm1_19}), and the fact that $\mu_j H(U_j|X_j)+(n-\mu_j)H(U_j)$ is minimized by $\mu_j=\mu$, which occurs when $S=\{(k,j): k\in\setS_p\}$, i.e., when the wiretapper observes the symbols of the $j$th transmitter in all the positions she chooses.

Using Hoeffding inequality and the definition of $\setD_{\gamma_j}^S$ in (\ref{eq:lemma2_2}), we have, for all $S\in\setS$,
\begin{align}
\label{eq:proof_Thm1_24}
&\Prob_{p_{\bU_j\bZ_S}}\left((\bU_j,\bZ_S)\notin\setD_{\gamma_j}^S\right)=\Prob_{p_{\bU_j\bZ_S}}\left(\log\frac{1}{p_{\bU_j|\bZ_S}(\bU_j|\bZ_S)}\leq \gamma_j\right)\\
\label{eq:proof_Thm1_25}
&=\Prob_{p_{\bU_j\bZ_S}}\left(\sum_{k\in \overline{S}_j}\log\frac{1}{p(U_{j,k}|X_{j,k})}+\sum_{k\in \overline{S}_j^c}\log\frac{1}{p(U_{j,k})}\leq (1-\tilde{\epsilon}_j)[\mu H(U_j|X_j)+(n-\mu )H(U_j)]\right)\\
\label{eq:proof_Thm1_26}
&\leq \Prob_{p_{\bU_j\bZ_S}}\left(\sum_{k\in \overline{S}_j}\log\frac{1}{p(U_{j,k}|X_{j,k})}+\sum_{k\in \overline{S}_j^c}\log\frac{1}{p(U_{j,k})}\leq (1-\tilde{\epsilon}_j)[\mu_jH(U_j|X_j)+(n-\mu_j)H(U_j)]\right)\\
\label{eq:proof_Thm1_27}
&\leq \exp(-\tilde{\beta}_j n),
\end{align}
where $\tilde{\beta}_j>0$ for $j=1,2$, and (\ref{eq:proof_Thm1_26}) follows because, for all $S\in\setS$,
\begin{align}
\label{eq:proof_Thm1_28}
\mu H(U_j|X_j)+(n-\mu )H(U_j)\leq \mu_j H(U_j|X_j)+(n-\mu_j )H(U_j).
\end{align}

Note that, for any finite $\gamma_j$, in order to compute the probability on the left hand side of (\ref{eq:proof_Thm1_24}), we only need to consider the tuples $(\bu_j,\bz)$ with $p_{\bU_j|\bZ_S}(\bu_j|\bz)>0$.
 
Similarly, for $i,j=1,2, i\neq j$ and all $S\in\setS$, using Hoeffding's inequality, (\ref{eq:proof_Thm1_21}), (\ref{eq:proof_Thm1_23}), and the definition for $\setD_{\gamma_{ij}}^S$ in (\ref{eq:lemma2_3}), we have 
\begin{align}
\label{eq:proof_Thm1_29}
&\Prob_{p_{\bU_{[1:2]}\bZ_S}}\left((\bU_{[1:2]},\bZ_S)\notin\setD_{\gamma_{ij}}^S\right)=\Prob_{p_{\bU_{[1:2]}\bZ_S}}\left(\log\frac{1}{p_{\bU_i|\bU_j\bZ_S}(\bU_i|\bU_j,\bZ_S)}\leq \gamma_{ij}\right)\leq \exp(-\tilde{\beta}_i n).
\end{align}

Taking $\delta^2=2\exp(-\tilde{\beta}n)$, where $\tilde{\beta}=\min\{\tilde{\beta}_1,\tilde{\beta}_2\}$, yields 
\begin{align}
\label{eq:proof_Thm1_30}
\Prob_{p_{\bU_{[1:2]}\bZ_S}}\left((\bU_{[1:2]},\bZ_S)\not\in\setD_{j}^S\right)\leq \delta^2,
\end{align}
for $j=1,2$ and all $S\in\setS$. Note that $\limitn\delta^2=0$, and hence, for $n$ sufficiently large, $\delta^2\in(0,\frac{1}{4})$. Thus, the conditions for Lemma \ref{lemma2} are satisfied. We also have, for $j=1,2,$ that
\begin{align}
\label{eq:proof_Thm1_31}
&\limitn(\delta+\delta^2)\log(\tW_j\tF_j)=\limitn n(R_j+\tilde{R}_j)(2\exp(-\tilde{\beta}n)+4\exp(-2\tilde{\beta}n))=0\\
\label{eq:proof_Thm1_32}
&\limitn H_b(\delta^2)=H_b\left(\limitn\delta^2\right)=0,
\end{align}
where (\ref{eq:proof_Thm1_32}) follows because $H_b$ is a continuous function. Thus, we have
\begin{align}
\label{eq:proof_Thm1_33}
&\limitn\tilde{\epsilon}=\epsilon+\limitn(\delta+\delta^2)\log(\tW_j\tF_j)+\limitn H_b(\delta^2)=\epsilon.
\end{align}

By substituting the choices for $\tW_j,\tF_j,\gamma_j,\gamma_{ij}$, where $i,j=1,2,\;i\neq j$, and 
\begin{align}
|\setS||\setZ^n|\leq \exp(n[(1+\alpha)\ln 2+\ln(|\setX_1|+|\setX_2|+1)]),
\end{align}
in (\ref{eq:lemma2}), and using (\ref{eq:proof_Thm1_33}), we have, for every $\epsilon,\epsilon'>0$, $\tilde{\epsilon}=\epsilon+\epsilon'$, there exist $n^*\in\mathbb{N}$ and $\kappa_\epsilon,\tilde{\kappa}>0$ such that for all $n\geq n^*$,
\begin{align}
\label{eq:proof_Thm1_34}
&\Prob_{\setB}\left(\underset{S\in\setS}\max\;\D\left(\tP_{W_{[1:2]}F_{[1:2]}\bZ_S}||p_{W_{[1:2]}}^Up_{F_{[1:2]}}^Up_{\bZ_S}\right)\geq 2\tilde{\epsilon}\right)\leq \exp\left(-\kappa_{\epsilon}e^{\tilde{\kappa} n}\right),
\end{align}
as long as 
\begin{align}
\label{eq:proof_Thm1_35}
&R_1+\tilde{R}_1\leq (1-\tilde{\epsilon}_1)\;[\alpha H(U_1|X_1)+(1-\alpha)H(U_1)],\\
\label{eq:proof_Thm1_36}
&R_2+\tilde{R}_2 \leq (1-\tilde{\epsilon}_2)\;[\alpha H(U_2|X_2)+(1-\alpha)H(U_2)].
\end{align}

By applying the first Borel-Cantelli Lemma to (\ref{eq:proof_Thm1_34}), we get
\begin{align}
\label{eq:proof_Thm1_38}
\limitn\Prob_{\setB}\left(\underset{S\in\setS}\max\;\D\left(\tP_{W_{[1:2]}F_{[1:2]}\bZ_S}||p_{W_{[1:2]}}^Up_{F_{[1:2]}}^Up_{\bZ_S}\right)>0\right)=0.
\end{align}

In addition, using Markov's inequality and (\ref{eq:proof_Thm1_5}), we have, for any $r>0$, that
\begin{align}
\label{eq:proof_Thm1_40}
\sum_{n=1}^\infty \Prob_{\setB}&\left(\V\left(\tP_{W_{[1:2]}F_{[1:2]}},p_{W_{[1:2]}}^Up_{F_{[1:2]}}^U\right)>r\right)\leq \frac{4}{r}\sum_{n=1}^\infty \exp(-\beta n)<\infty.
\end{align}
Using the first Borel-Cantelli lemma, it follows from (\ref{eq:proof_Thm1_40}) that  
\begin{align}
\label{eq:proof_Thm1_41}
\limitn\Prob_{\setB}&\left(\V\left(\tP_{W_{[1:2]}F_{[1:2]}},p_{W_{[1:2]}}^Up_{F_{[1:2]}}^U\right)>0 \right)=0.
\end{align}

\begin{remark}
{\emph{In the secrecy condition for the source model, (\ref{eq:proof_Thm1_38}), we require the independence of the public messages $F_{[1:2]}$ from the confidential keys $W_{[1:2]}$ and the wiretapper's observation $\bZ_S$. The reason is that, after showing that the secrecy condition in (\ref{eq:proof_Thm1_38}) holds as well for the channel model in protocol B, we need to eliminate the common randomness $F_{[1:2]}$ from the channel model by conditioning on a certain instance of it, without distributing the established independence between the messages $W_{[1:2]}$ and the wiretapper's observation $\bZ_S$.}}
\end{remark}

Now, we show that the reliability and secrecy conditions in (\ref{eq:proof_Thm1_11}) and (\ref{eq:proof_Thm1_38}) hold as well for the channel model in protocol B. First, for the reliability condition, using (\ref{eq:P_tilde_2}), (\ref{eq:P_1}), and the triangle inequality, we have 
\begin{align}
\label{eq:proof_Thm1_42}
\nonumber &\V\left(P_{W_{[1:2]}F_{[1:2]}\bU_{[1:2]}\bY\bZ_S\bUh_{[1:2]}},P_{W_{[1:2]}F_{[1:2]}\bU_{[1:2]}\bY\bZ_S}\mathbbm{1}\{\bUh_{[1:2]}=\bU_{[1:2]}\}\right)\\
\nonumber &\leq \V\left(P_{W_{[1:2]}F_{[1:2]}\bU_{[1:2]}\bY\bZ_S\bUh_{[1:2]}},\tP_{W_{[1:2]}F_{[1:2]}\bU_{[1:2]}\bY\bZ_S\bUh_{[1:2]}}\right)\\
\nonumber &\qquad +\V\left(\tP_{W_{[1:2]}F_{[1:2]}\bU_{[1:2]}\bY\bZ_S\bUh_{[1:2]}},\tP_{W_{[1:2]}F_{[1:2]}\bU_{[1:2]}\bY\bZ_S}\mathbbm{1}\{\bUh_{[1:2]}=\bU_{[1:2]}\}\right)\\
&\qquad +\V\left(\tP_{W_{[1:2]}F_{[1:2]}\bU_{[1:2]}\bY\bZ_S}\mathbbm{1}\{\bUh_{[1:2]}=\bU_{[1:2]}\},P_{W_{[1:2]}F_{[1:2]}\bU_{[1:2]}\bY\bZ_S}\mathbbm{1}\{\bUh_{[1:2]}=\bU_{[1:2]}\}\right)\\
\label{eq:proof_Thm1_43}
\nonumber &= \V\left(\tP_{W_{[1:2]}F_{[1:2]}\bU_{[1:2]}\bY\bZ_S\bUh_{[1:2]}},\tP_{W_{[1:2]}F_{[1:2]}\bU_{[1:2]}\bY\bZ_S}\mathbbm{1}\{\bUh_{[1:2]}=\bU_{[1:2]}\}\right)\\
&\qquad\qquad\qquad\quad +2 \V\left(P_{W_{[1:2]}F_{[1:2]}},p_{W_{[1:2]}}^U p_{F_{[1:2]}}^U\right).
\end{align}
Thus, using (\ref{eq:proof_Thm1_5}), (\ref{eq:proof_Thm1_11}), and (\ref{eq:proof_Thm1_43}), we have 
\begin{align}
\label{eq:proof_Thm1_44}
&\limitn\E_{\setB}\left(\V\left(P_{W_{[1:2]}F_{[1:2]}\bU_{[1:2]}\bY\bZ_S\bUh_{[1:2]}},P_{W_{[1:2]}F_{[1:2]}\bU_{[1:2]}\bY\bZ_S}\mathbbm{1}\{\bUh_{[1:2]}=\bU_{[1:2]}\}\right)\right)=0.
\end{align}
 
Second, for the secrecy condition, using the union bound, we have
\begin{align}
\nonumber&\Prob_{\setB}\left(\max_{S\in\setS} \;\D\left(P_{W_{[1:2]}F_{[1:2]}\bZ_S}||p_{W_{[1:2]}}^Up_{F_{[1:2]}}^Up_{\bZ_S}\right)>0\right)\\ 
\label{eq:proof_Thm1_46}
&\leq \Prob_{\setB}\left(\max_{S\in\setS}\;\D\left(\tP_{W_{[1:2]}F_{[1:2]}\bZ_S}||p_{W_{[1:2]}}^Up_{F_{[1:2]}}^Up_{\bZ_S}\right)>0\right)+\Prob_{\setB} \left(\V\left(\tP_{W_{[1:2]}F_{[1:2]}},p_{W_{[1:2]}}^Up_{F_{[1:2]}}^U\right)>0\right).
\end{align}
Thus, using (\ref{eq:proof_Thm1_38}), (\ref{eq:proof_Thm1_41}),  and (\ref{eq:proof_Thm1_46}), we have 
\begin{align}
\label{eq:proof_Thm1_47}
\limitn\Prob_{\setB}\left(\max_{S\in\setS} \;\D\left(P_{W_{[1:2]}F_{[1:2]}\bZ_S}||p_{W_{[1:2]}}^Up_{F_{[1:2]}}^Up_{\bZ_S}\right)>0\right)=0.
\end{align}

By applying the selection lemma to (\ref{eq:proof_Thm1_44}) and (\ref{eq:proof_Thm1_47}), there is at least one binning realization $\bold{b}^*=\{b_1^{*(j)},b_2^{*(j)}:j=1,2\}$, with a corresponding joint distribution $p^*$ for protocol B such that  
\begin{align}
\label{eq:proof_Thm1_48}
&\limitn\V\left(p^*_{W_{[1:2]}F_{[1:2]}\bU_{[1:2]}\bY\bZ_S\bUh_{[1:2]}}, p^*_{W_{[1:2]}F_{[1:2]}\bU_{[1:2]}\bY\bZ_S}\mathbbm{1}\{\bUh_{[1:2]}=\bU_{[1:2]}\}\right)= 0,\\
\label{eq:proof_Thm1_49}
&\text{and }\;\limitn\mathbbm{1}\left\{\max_{S\in\setS} \;\D\left(p^*_{W_{[1:2]}F_{[1:2]}\bZ_S}||p_{W_{[1:2]}}^Up_{F_{[1:2]}}^Up_{\bZ_S}\right)>0\right\}=0,
\end{align}
where $W_j=b_1^{*(j)}(\bU_j)$ and $F_j=b_2^{*(j)}(\bU_j)$, $j=1,2$. 

Next, we introduce the $\hat{W}$ variables to the joint distributions in (\ref{eq:proof_Thm1_48}). For $j=1,2,$ $\hat{W}_j$ is a deterministic function of the random sequence $\bUh_j$. In particular, $p^*_{\hat{W}_j|\bUh_j}(\hat{w}_j|\buh_j)=\mathbbm{1}\left\{\hat{w}_j=b_1^{*(j)}(\buh)\right\}$. Using (\ref{eq:proof_Thm1_48}) and a similar analysis as in \cite[($58$)-($64$)]{nafea2017new}, we have
\begin{align}
\label{eq:proof_Thm1_50}
\nonumber\limitn&\E_{F_{[1:2]}}\left(\Prob_{p^*}\left(\hat{W}_{[1:2]}\neq W_{[1:2]}|F_{[1:2]}\right)\right)\\
&=\limitn\V\left(p^*_{W_{[1:2]}F_{[1:2]}\bU_{[1:2]}\bY\bZ_S\bUh_{[1:2]}}, p^*_{W_{[1:2]}F_{[1:2]}\bU_{[1:2]}\bY\bZ_S}\mathbbm{1}\{\bUh_{[1:2]}=\bU_{[1:2]}\}\right)= 0.
\end{align}

Using the union bound, we also have 
\begin{align}
\nonumber& \Prob_{F_{[1:2]}}\left(\max_{S\in\setS}\;\D\left(p^*_{W_{[1:2]}\bZ_S|F_{[1:2]}}||p_{W_{[1:2]}}^Up^*_{\bZ_S|F_{[1:2]}}\right)>0\right)\\
\label{eq:proof_Thm1_52}
\nonumber&\qquad =\Prob\left(\max_{S\in\setS}\D(p^*_{W_{[1:2]}\bZ_S|F_{[1:2]}}||p_{W_{[1:2]}}^Up^*_{\bZ_S|F_{[1:2]}})>0,\text{ and }\forall S,\; p^*_{W_{[1:2]}F_{[1:2]}\bZ_S}=p_{W_{[1:2]}}^Up_{F_{[1:2]}}^Up_{\bZ_S}\right)\\
&\qquad\qquad\qquad +\mathbbm{1}\left\{\max_{S\in\setS} \D\left(p^*_{W_{[1:2]}F_{[1:2]}\bZ_S}||p_{W_{[1:2]}}^Up_{F_{[1:2]}}^Up_{\bZ_S}\right)>0\right\}\\
\label{eq:proof_Thm1_53}
&\qquad =\mathbbm{1}\left\{\max_{S\in\setS} \D\left(p^*_{W_{[1:2]}F_{[1:2]}\bZ_S}||p_{W_{[1:2]}}^Up_{F_{[1:2]}}^Up_{\bZ_S}\right)>0\right\},
\end{align}
where (\ref{eq:proof_Thm1_53}) follows since the first term on the right hand side of (\ref{eq:proof_Thm1_52}) is equal to zero. 
Thus, using (\ref{eq:proof_Thm1_49}) and (\ref{eq:proof_Thm1_53}), we have 
\begin{align}
\label{eq:proof_Thm1_54}
\limitn\Prob_{F_{[1:2]}}\left(\max_{S\in\setS}\;\D\left(p^*_{W_{[1:2]}\bZ_S|F_{[1:2]}}||p_{W_{[1:2]}}^Up^*_{\bZ_S|F_{[1:2]}}\right)>0\right)=0
\end{align}

Once again, applying the selection lemma to (\ref{eq:proof_Thm1_50}) and (\ref{eq:proof_Thm1_54}), implies that there is at least one realization  $f_{[1:2]}^{*}$ such that 
\begin{align}
&\limitn \Prob\left(\hat{W}_{[1:2]}\neq W_{[1:2]}|F_{[1:2]}=f_{[1:2]}^*\right)=0,\\
&\limitn \max_{S\in\setS} I\left(W_{[1:2]};\bZ_S|F_{[1:2]}=f_{[1:2]}^{*}\right)=0.
\end{align}
Let $\tilde{p}^{*}$ be the induced joint distribution for protocol A which corresponds to the binning realization $\bold{b}^*$. We identify $\left\{\tilde{p}^{*}(\bu_j|w_j,f^*_j),p(\bx_j|\bu_j),j=1,2\right\}$ and $\left\{\tilde{p}^*(\buh_{[1:2]}|\by,f^*_{[1:2]}), \{b_1^{*(j)}(\buh_j),j=1,2\}\right\}$ as the encoders and the decoder for the original channel model. 

By combining the rate conditions in (\ref{eq:proof_Thm1_3}), (\ref{eq:proof_Thm1_4}), (\ref{eq:proof_Thm1_8})-(\ref{eq:proof_Thm1_10}), (\ref{eq:proof_Thm1_35}), and (\ref{eq:proof_Thm1_36}), and taking $\tilde{\epsilon}_1,\tilde{\epsilon}_2\rightarrow 0$, we obtain the achievable strong secrecy rate region in (\ref{eq:Thm1_1})-(\ref{eq:Thm1_3}). The convex hull follows by time sharing independent codes and the fact that maximizing the secrecy constraint over $S$ in the whole block-length is upper bounded by its maximization over the individual segments of the time sharing. 

\section{Proofs for Theorems \ref{thm:Thm2} and \ref{thm:Thm3}}{\label{Achievability_Thm2_3}}
The proof for Theorem \ref{thm:Thm2} follows similar steps as in the proof for Theorem \ref{thm:Thm1}. The difference is that $\setS$ and $\bZ_S$, for all $S\in\setS$, in protocol A are defined as in (\ref{eq:setS_Model_B}), (\ref{eq:Z_S_n_B}). We thus have, for $i,j=1,2,\;i\neq j$, and all $S\in\setS$, that
\begin{align}
\label{eq:proof_Thm2_1}
&H(\bU_j|\bZ_S)=H(\bU_{j,S},\bU_{j,S^c}|\bX_{1,S}+\bX_{2,S})\\
\label{eq:proof_Thm2_2}
&\qquad =H(\bU_{j,S}|\bX_{1,S}+\bX_{2,S})+H(\bU_{j,S^c}|\bU_{j,S},\bX_{1,S}+\bX_{2,S})\\
\label{eq:proof_Thm2_3}
&\qquad =H(\bU_{j,S}|\bX_{1,S}+\bX_{2,S})+H(\bU_{j,S^c})\\
\label{eq:proof_Thm2_4}
&\qquad =\mu H(U_j|X_1+X_2)+(n-\mu)H(U_j)\\
\label{eq:proof_Thm2_5}
&H(\bU_i|\bU_j,\bZ_S)=H(\bU_{i,S},\bU_{i,S^c}|\bU_j,\bX_{1,S}+\bX_{2,S})\\
\label{eq:proof_Thm2_6}
&\qquad=H(\bU_{i,S}|\bU_j,\bX_{1,S}+\bX_{2,S})+H(\bU_{i,S^c}|\bU_{i,S},\bU_j,\bX_{1,S}+\bX_{2,S})\\
\label{eq:proof_Thm2_7}
&\qquad=H(\bU_{i,S}|\bU_{j,S},\bX_{1,S}+\bX_{2,S})+H(\bU_{i,S^c})\\
\label{eq:proof_Thm2_8}
&\qquad=\mu H(U_i|U_j,X_1+X_2)+(n-\mu)H(U_i).
\end{align} 

Thus, in applying Lemma \ref{lemma2} to the source model in protocol A, for $i,j=1,2,\;i\neq j$, and $\tilde{\epsilon}_j>0$, we choose 
\begin{align}
\label{eq:proof_Thm2_9}
&\gamma_j=(1-\tilde{\epsilon}_j)\min_{S\in\setS}\;H(\bU_j|\bZ_S)=(1-\tilde{\epsilon}_j)[\mu H(U_j|X_1+X_2)+(n-\mu)H(U_j)]\\
\label{eq:proof_Thm2_10}
&\gamma_{ij}=(1-\tilde{\epsilon}_j)\min_{S\in\setS}\;H(\bU_i|\bU_j,\bZ_S)=(1-\tilde{\epsilon}_j)[\mu H(U_i|U_j,X_1+X_2)+(n-\mu)H(U_i)].
\end{align}
Using Hoeffding inequality, the conditions of the lemma are satisfied, and the rate conditions required for the secrecy property in (\ref{eq:proof_Thm1_38}) are 
\begin{align}
\label{eq:proof_Thm2_11}
&R_1+\tilde{R}_1\leq \alpha H(U_1|X_1+X_2)+(1-\alpha)H(U_1)\\
\label{eq:proof_Thm2_11}
&R_2+\tilde{R}_2\leq \alpha H(U_2|X_1+X_2)+(1-\alpha)H(U_2)\\
\label{eq:proof_Thm2_13}
&R_1+R_2+\tilde{R}_1+\tilde{R}_2\leq \alpha H(U_{[1:2]}|X_1+X_2)+(1-\alpha)H(U_{[1:2]}).
\end{align}   
These conditions, combined with the rate conditions for the Slepian-Wolf decoder, which are 
\begin{align}
\label{eq:proof_Thm2_14}
&\tilde{R}_1\geq H(U_1|U_2,Y),\qquad \tilde{R}_2\geq H(U_2|U_1,Y),\\
\label{eq:proof_Thm2_15}
&\tilde{R}_1+\tilde{R}_2\geq H(U_{[1:2]}|Y),
\end{align}
and using time sharing, establish the achievability for the strong secrecy rate region in Theorem \ref{thm:Thm2}.

\begin{remark}
{\emph{By setting $j=1,i=2$, instead of the minimum in the right hand side of (\ref{eq:lemma2}), Lemma 2 results in the maximum binning rate
$R_1+\tilde{R}_1$ of the source $\bU_1$, and the corresponding maximum conditional binning rate $R_2+\tilde{R}_2$ for the source $\bU_2$ given $R_1+\tilde{R}_1$, such that the probability in the left hand side 
of (\ref{eq:lemma2}) is vanishing. In other words, Lemma \ref{lemma2} provides the corner points of the binning rate region such that the probability, over the random binning of the sources, that the bins are independent, uniform, and independent from the wiretapper's observation, is vanishing.}}
\end{remark}

Similarly, the proof for Theorem \ref{thm:Thm3} follows similar steps as in the proof for Theorem \ref{thm:Thm1}. In protocol A, $\setS$ and $\bZ_S$ for all $S\in\setS$ are defined as in (\ref{eq:Z_S_n}) in Section \ref{New_MAC_WTC}. The sequences $\bU_1,\bU_2$ are i.i.d. and the channel $p_{V|U_{[1:2]}}$ is a discrete memoryless channel, since it results from concatenating the two discrete memoryless channels $p_{V|X_{[1:2]}}$ and $p_{X_{[1:2]}|U_{[1:2]}}$. Thus, we have, for $i,j=1,2,\;i\neq j,$ and all $S\in\setS,$
\begin{align}
\label{eq:proof_Thm3_1}
&H(\bU_j|\bZ_S)=H(\bU_{j,S},\bU_{j,S^c}|\bX_{1,S},\bX_{2,S},\bV_{S^c})\\
\label{eq:proof_Thm3_2}
&\qquad =H(\bU_{j,S}|\bX_{1,S},\bX_{2,S},\bV_{S^c})+H(\bU_{j,S^c}|\bU_{j,S},\bX_{1,S},\bX_{2,S},\bV_{S^c})\\
\label{eq:proof_Thm3_3}
&\qquad =H(\bU_{j,S}|\bX_{j,S})+H(\bU_{j,S^c}|\bV_{S^c})\\
\label{eq:proof_Thm3_4}
&\qquad = \mu H(U_j|X_j)+(n-\mu) H(U_j|V)\\
\label{eq:proof_Thm3_5}
&H(\bU_i|\bU_j,\bZ_S)=H(\bU_{i,S},\bU_{i,S^c}|\bU_j,\bX_{1,S},\bX_{2,S},\bV_{S^c})\\
\label{eq:proof_Thm3_6}
&\qquad=H(\bU_{i,S}|\bU_j,\bX_{1,S},\bX_{2,S},\bV_{S^c})+H(\bU_{i,S^c}|\bU_{i,S},\bU_{j,S},\bU_{j,S^c},\bX_{1,S},\bX_{2,S},\bV_{S^c})\\
\label{eq:proof_Thm3_7}
&\qquad=H(\bU_{i,S}|\bX_{i,S})+H(\bU_{i,S^c}|\bU_{j,S^c},\bV_{S^c})\\
\label{eq:proof_Thm3_8}
&\qquad=\mu H(U_i|X_i)+(n-\mu )H(U_i|U_j,V),
\end{align}
where (\ref{eq:proof_Thm3_3}) follows due to the Markov chains $\bU_{j,S}-\bX_{j,S}-(\bX_{i,S},\bV_{S^c})$ and $(\bU_{j,S},\bX_{1,S},\bX_{2,S})-\bV_{S^c}-\bU_{j,S^c}$. Equation (\ref{eq:proof_Thm3_7}) follows from the Markov chains $\bU_{i,S}-\bX_{i,S}-(\bU_j,\bX_{j,S},\bV_{S^c})$ and $(\bU_{i,S},\bU_{j,S},\bX_{1,S},\bX_{2,S})-(\bU_{j,S^c},\bV_{S^c})-\bU_{i,S^c}$. These Markov chains follow since the sequences $\bU_1,\bU_2$ are i.i.d. and the channels $p_{X_1|U_1},p_{X_2|U_2}, p_{V|U_{[1:2]}}$ are discrete memoryless. 

Thus, for $i,j=1,2,\;i\neq j$, and $\tilde{\epsilon}_j>0$, by choosing 
\begin{align}
\label{eq:proof_Thm3_9}
&\gamma_j=(1-\tilde{\epsilon}_j)\min_{S\in\setS}\;H(\bU_j|\bZ_S)=(1-\tilde{\epsilon}_j)[\mu H(U_j|X_j)+(n-\mu) H(U_j|V)]\\
\label{eq:proof_Thm3_10}
&\gamma_{ij}=(1-\tilde{\epsilon}_j)\min_{S\in\setS}\;H(\bU_i|\bU_j,\bZ_S)=(1-\tilde{\epsilon}_j)[\mu H(U_i|X_i)+(n-\mu )H(U_i|U_j,V)],
\end{align}
and using Hoeffding inequality, the conditions of Lemma \ref{lemma2} are satisfied. The rate conditions needed for the secrecy property in (\ref{eq:proof_Thm1_38}) are 
\begin{align}
\label{eq:proof_Thm3_11}
&R_1+\tilde{R}_1\leq \alpha H(U_1|X_1)+(1-\alpha)H(U_1|V)\\
\label{eq:proof_Thm3_12}
&R_2+\tilde{R}_2\leq \alpha H(U_2|X_2)+(1-\alpha)H(U_2|V)\\
\label{eq:proof_Thm3_13}
&R_1+R_2+\tilde{R}_1+\tilde{R}_2\leq \alpha H(U_{[1:2]}|X_{[1:2]})+(1-\alpha)H(U_{[1:2]}|V).
\end{align}   
Combining (\ref{eq:proof_Thm3_11})-(\ref{eq:proof_Thm3_13}) with the rate conditions required for the Slepian-Wolf decoder in (\ref{eq:proof_Thm2_14}) and (\ref{eq:proof_Thm2_15}), and using time sharing, establish the achievability for the strong secrecy rate region in Theorem \ref{thm:Thm3}.

\section{Conclusion}\label{Con}
In this paper, we have studied the extension of the wiretap channel II with a noisy main channel in \cite{nafea2015wiretap} and the generalized wiretap channel model in \cite{nafea2017new} to the multiple access setting. For the multiple access wiretap channel II with a noisy main channel, we have proposed three attack models for the wiretapper and derived an achievable strong secrecy rate region for each. We have generalized the strongest attack model, in which the wiretapper observes the transmitted symbols of both users in the positions of the subset she chooses, to the case when the wiretapper observes the outputs of a noisy multiple access channel instead of erasures outside this subset, proposing a {\it{generalized}} multiple access wiretap model. We have derived an achievable strong secrecy rate region for this generalized  model. This model generalizes the multiple access wiretap channel in \cite{tekin2008gaussian,tekin2008general} as well to the case when the wiretapper is provided with noiseless observations for a subset, of her choice, of the transmitted codeword symbols of both uses. The tools we have utilized for achievability extend the set of tools utilized for the single-user scenario in \cite{nafea2017new} to a multi-user setting. Future work includes other multi-terminal setups with more capable wiretappers. 
%%------------------------------------------------------------------------------------------------

\appendix
First, we rewrite the relative entropy in (\ref{eq:lemma2}) as follows:
\begin{align}
\label{eq:appendix_B_1}
\nonumber &\D\left(P_{W_{[1:2]}F_{[1:2]}Z_S}||p_{W_{[1:2]}}^Up_{F_{[1:2]}}^U p_{Z_S}\right)\\
&=\sum_{w_{[1:2]},f_{[1:2]},z}P_{W_{[1:2]}F_{[1:2]}Z_S}(w_{[1:2]},f_{[1:2]},z)\log\frac{P_{W_{[1:2]}F_{[1:2]}Z_S}(w_{[1:2]},f_{[1:2]},z)}{p_{W_{[1:2]}}^Up_{F_{[1:2]}}^U p_{Z_S}(z)}\\
\label{eq:appendix_B_2}
&=\sum_{w_{[1:2]},f_{[1:2]},z}P_{W_{[1:2]}F_{[1:2]}Z_S}(w_{[1:2]},f_{[1:2]},z)\log\left(\frac{P_{W_{[1:2]}F_{[1:2]} Z_S}(w_{[1:2]},f_{[1:2]},z)}{P_{W_1F_1 Z_S}(w_1,f_1,z) p_{W_2}^U p_{F_2}^U}.\frac{P_{W_1F_1Z_S}(w_1,f_1,z)}{p_{W_1}^U p_{F_1}^Up_{Z_S}(z)}\right)\\
\label{eq:appendix_B_3}
&=\E_{p_{Z_S}}\left(\D\left(P_{W_{[1:2]}F_{[1:2]}|Z_S}||P_{W_1F_1|Z_S}p_{W_2}^Up_{F_2}^U\right)\right)+\D\left(P_{W_1F_1Z_S}||p_{W_1}^Up_{F_1}^Up_{Z_S}\right).
\end{align}
Thus, the probability in (\ref{eq:lemma2}) is upper bounded as 
\begin{align}
\label{eq:appendix_B_4}
\nonumber&\Prob_{\setB}\left(\underset{S\in\setS}\max\; \D\left(P_{W_{[1:2]}F_{[1:2]}Z_S}||p_{W_{[1:2]}}^Up_{F_{[1:2]}}^U p_{Z_S}\right)\geq 2\tilde{\epsilon}\right)\leq \Prob_{\setB}\left(\max_{S\in\setS}\D\left(P_{W_1F_1Z_S}||p_{W_1}^Up_{F_1}^Up_{Z_S}\right)>\tilde{\epsilon}\right)\\
&\qquad\quad +\Prob_{\setB}\left(\max_{S\in\setS}\E_{p_{Z_S}}\D\left(P_{W_{[1:2]}F_{[1:2]}|Z_S}||P_{W_1F_1|Z_S}p_{W_2}^Up_{F_2}^U\right)>\tilde{\epsilon}\right).
\end{align}

We upper bound each term on the right hand side of (\ref{eq:appendix_B_4}). Using \cite[Lemma 2]{nafea2017new}, the first term is upper bounded as 
\begin{align}
\label{eq:appendix_B_61}
\Prob_{\setB}\left(\max_{S\in\setS}\D\left(P_{W_1F_1Z_S}||p_{W_1}^Up_{F_1}^Up_{Z_S}\right)>\tilde{\epsilon}\right)\leq |\setS||\setZ|\exp\left(\frac{-\epsilon^2 (1-\delta) 2^{\gamma_1}}{3\tW_1\tF_1}\right).
\end{align}

Next, we upper bound the second term in (\ref{eq:appendix_B_4}). For all $S\in\setS$, let us define 
\begin{align}
\label{eq:appendix_B_5_0}
\setA_S\triangleq\left\{z\in\setZ:\Prob_{p_{X_{[1:2]}|Z_S}}\left((X_{[1:2]},z)\in \setD_{1}^S\right)\geq 1-\delta\right\},
\end{align}
where $\setD_{1}^S$ is defined in (\ref{eq:lemma2_1}). We have
\begin{align}
\label{eq:appendix_B_5}
\Prob_{p_{Z_S}}(\setA_S^c)&=\Prob_{p_{Z_S}}\left(\Prob_{p_{X_{[1:2]}|Z_S}}\left((X_{[1:2]},z)\notin \setD_{1}^S\right)\geq \delta\right)\\
\label{eq:appendix_B_6}
&\leq\frac{1}{\delta}\;\E_{p_{Z_S}}\left(\Prob_{p_{X_{[1:2]}|Z_S}}\left((X_{[1:2]},z)\notin \setD_{1}^S\right)\right)\\
\label{eq:appendix_B_8}
&=\frac{1}{\delta}\;\Prob_{p_{X_{[1:2]}Z_S}}\big((X_{[1:2]},Z_S)\notin\setD_1^S\big)\\
\label{eq:appendix_B_9}
&\leq \frac{\delta^2}{\delta}=\delta.
\end{align}
where (\ref{eq:appendix_B_6}) follows from Markov's inequality, and the inequality in (\ref{eq:appendix_B_9}) follows from (\ref{eq:lemma2_4}).

For all $w_{[1:2]},f_{[1:2]}\in [1:\tilde{W}]\times[1:\tilde{F}]$ , $z\in\setZ$, and $S\in\setS$, define
\begin{align}
\label{eq:appendix_B_10}
\nonumber {P}_1^S(w_{[1:2]},f_{[1:2]}|z)=\sum_{x_{[1:2]}\in\setX_1\times\setX_2} &p_{X_{[1:2]}|Z_S}(x_{[1:2]}|z)\mathbbm{1}\left\{(x_{[1:2]},z)\in\setD_{1}^S\right\}\\
&\times\mathbbm{1}\left\{\setB_1^{(j)}(x_j)=w_j,\setB_2^{(j)}(x_j)=f_j,\forall j=1,2\right\}\\
\label{eq:appendix_B_11}
\nonumber {P}_2^S(w_{[1:2]},f_{[1:2]}|z)=\sum_{x_{[1:2]}\in\setX_1\times\setX_2} &p_{X_{[1:2]|Z_S}}(x_{[1:2]}|z)\mathbbm{1}\left\{(x_{[1:2]},z)\notin\setD_{1}^S\right\}\\
&\times\mathbbm{1}\left\{\setB_1^{(j)}(x_j)=w_j,\setB_2^{(j)}(x_j)=f_j,\forall j=1,2\right\}.
\end{align}
Thus, we have $P_{W_{[1:2]}F_{[1:2]}|Z_S}(w_{[1:2]},f_{[1:2]}|z)=P_1^S(w_{[1:2]},f_{[1:2]}|z)+P_2^S(w_{[1:2]},f_{[1:2]}|z)$. 

Now, for every $x_2\in\setX_2$, define
\begin{align}
\label{eq:appendix_B_12}
U_{x_2}=\sum_{x_1\in\setX_1}p_{X_{[1:2]}|Z_S}(x_{[1:2]}|z)\mathbbm{1}\left\{\setB_1^{(2)}(x_2)=w_2,\setB_2^{(2)}(x_2)=f_2\right\}\mathbbm{1}\left\{(x_{[1:2]},z)\in \setD_{1}^S\right\}.
\end{align}  
The random variables $\left\{U_{x_2}\right\}_{x_2\in\setX_2}$ are non-negative and independent since the random variables $\left\{\setB_1^{(2)}(x_2),\setB_2^{(2)}(x_2)\right\}_{x_2\in\setX_2}$ are independent. From the definition of $\setD_1^S$ in (\ref{eq:lemma2_1}), we have for $(x_{[1:2]},z)\in \setD_{1}^S$ that $(x_{[1:2]},z)\in\setD_{\gamma_{21}}^S$. Additionally, from the definition of $\setD_{\gamma_{21}}$ in (\ref{eq:lemma2_3}), we have that $p(x_2|x_1,z)\leq 2^{-\gamma_{21}}$. From (\ref{eq:appendix_B_12}), we have 
\begin{align}
\label{eq:appendix_B_13}
U_{x_2}&\leq \sum_{x_1}p_{X_1|Z_S}(x_1|z)p_{X_2|X_1,Z_S}(x_2|x_1,z)\mathbbm{1}\left\{(x_{[1:2]},z)\in \setD_{1}^S\right\}\\
\label{eq:appendix_B_14}
&\leq 2^{-\gamma_{21}}\sum_{x_1} p_{X_1|Z_S}(x_1|z)\mathbbm{1}\left\{(x_{[1:2]},z)\in \setD_{1}^S\right\}\\
\label{eq:appendix_B_15}
&\leq 2^{-\gamma_{21}}.
\end{align}

Since for all $x_2\in\setX_2$,
\begin{align}
\label{eq:appendix_B_16}
\E_{\setB}\left(\mathbbm{1}\left\{\setB_1^{(2)}(x_2)=w_2,\setB_2^{(2)}(x_2)=f_2\right\}\right)=\frac{1}{\tW_2\tF_2},
\end{align}
we have, 
\begin{align}
\label{eq:appendix_B_17}
\sum_{x_2\in\setX_2}\E_{\setB}(U_{x_2})&=\frac{1}{\tW_2\tF_2}\sum_{x_{[1:2]}\in\setX_1\times\setX_2}p_{X_{[1:2]}|Z_S}(x_{[1:2]}|z)\mathbbm{1}\left\{(x_{[1:2]},z)\in \setD_{1}^S\right\}\\
\label{eq:appendix_B_18}
&=\frac{\Prob_{p_{X_{[1:2]}|Z_S}}\left(\left(X_{[1:2]},z\right)\in\setD_1^S\right)}{\tW_2\tF_2}.
\end{align}

In addition, notice that 
\begin{align}
\label{eq:appendix_B_19}
\nonumber \sum_{w_1,f_1}&P_1^S(w_{[1:2]},f_{[1:2]}|z)=\sum_{x_{[1:2]}}p_{X_{[1:2]}|Z_S}(x_{[1:2]}|z)\mathbbm{1}\left\{\left(x_{[1:2]},z\right)\in\setD_1^S\right\}\\
&\qquad \qquad\qquad \qquad \times\sum_{w_1,f_1}\mathbbm{1}\left\{\setB_1^{(j)}(x_j)=w_j,\setB_2^{(j)}(x_j)=f_j,\forall j=1,2\right\}\\
\label{eq:appendix_B_20}
&=\sum_{x_2}\sum_{x_1}p_{X_{[1:2]}|Z_S}(x_{[1:2]}|z)\mathbbm{1}\left\{\setB_1^{(2)}(x_2)=w_2,\setB_2^{(2)}(x_2)=f_2\right\}\mathbbm{1}\left\{\left(x_{[1:2]},z\right)\in\setD_1^S\right\}\\
\label{eq:appendix_B_21_0}
&=\sum_{x_2}U_{x_2}
\end{align}

We now state the following lemma, which is a variation on Chernoff's bound that we need to utilize in the proof. 
\begin{lemma}\label{lemma_chernoff}(A variation on Chernoff bound \cite[Lemma 6]{nafea2017new}):
Let $U_1,U_2,\cdots,U_n$ be a sequence of non-negative independent random variables with respective means $\E(U_i)=\bar{m}_i$. If $U_i\in[0,b]$, for all $i\in[1:n]$, and $\sum_{i=1}^n\bar{m}_i\leq \bar{m}$, then, for every $ \epsilon\in[0,1]$, we have
\begin{align}
\label{eq:lemma_chernoff}
\Prob\left(\sum_{i=1}^n U_i\geq (1+\epsilon)\bar{m}\right)\leq \exp\left(-\epsilon^2\frac{\bar{m}}{3b}\right).
\end{align}
\end{lemma}

The random variables $\left\{U_{x_2}\right\}_{x_2\in\setX_2}$ are non-negative, independent, and $U_{x_2}\in[0,2^{-\gamma_{21}}]$ for all $x_2\in\setX_2$. By applying Lemma \ref{lemma_chernoff} to the random variables $\{U_{x_2}\}_{x_2\in\setX_2}$, we have, 
\begin{align}
\label{eq:appendix_B_21}
\nonumber \Prob_{\setB}&\left({P}_1^S(w_{[1:2]},f_{[1:2]}|z)\geq \frac{1+\epsilon}{\tW_2\tF_2}P_{W_1F_1|Z_S}(w_1,f_1|z)\right)\\
&\leq \Prob_{\setB}\left(\sum_{w_1,f_1}{P}_1^S(w_{[1:2]},f_{[1:2]}|z)\geq \frac{1+\epsilon}{\tW_2\tF_2}\sum_{w_1,f_1} P_{W_1F_1|Z_S}(w_1,f_1|z)\right)\\
\label{eq:appendix_B_22}
&=\Prob\left(\sum_{x_2} U_{x_2} \geq \frac{1+\epsilon}{\tW_2\tF_2}\right)\\
\label{eq:appendix_B_23}
&\leq\Prob\left(\sum_{x_2} U_{x_2} \geq \frac{1+\epsilon}{\tW_2\tF_2} \Prob_{p_{X_{[1:2]}|Z_S}}\left(\left(X_{[1:2]},z\right)\in\setD_1^S\right)\right)\\
\label{eq:appendix_B_24}
&=\Prob\left(\sum_{x_2} U_{x_2} \geq \left(1+\epsilon\right)\sum_{x_2}\E_{\setB}(U_{x_2}) \right)\\
\label{eq:appendix_B_25}
&\leq \exp{\left(\frac{-\epsilon^2 2^{\gamma_{21}}}{3\tW_2\tF_2} \Prob_{p_{X_{[1:2]}|Z_S}}\left(\left(X_{[1:2]},z\right)\in\setD_1^S\right)\right)}.
\end{align}
where (\ref{eq:appendix_B_22}) follows from (\ref{eq:appendix_B_21_0}), (\ref{eq:appendix_B_24}) follows from (\ref{eq:appendix_B_18}), and (\ref{eq:appendix_B_25}) follows from Lemma \ref{lemma_chernoff}. 

From the definition of $\setA_S$ in (\ref{eq:appendix_B_5_0}), we have, for all $z\in\setA_S$, that $\Prob_{p_{X_{[1:2]}|Z_S}}\left(\left(X_{[1:2]},z\right)\in\setD_1^S\right)\geq 1-\delta$. Thus, for all $z\in\setA_S$,
\begin{align}
\label{eq:appendix_B_26}
\Prob_{\setB}\left({P}_1^S(w_{[1:2]},f_{[1:2]}|z)\geq \frac{1+\epsilon}{\tW_2\tF_2}P_{W_1F_1|Z_S}(w_1,f_1|z)\right)\leq \exp{\left(\frac{-\epsilon^2 (1-\delta)2^{\gamma_{21}}}{3\tW_2\tF_2} \right)}.
\end{align}

Note that, for fixed $z\in\setZ$ and $S\in\setS$, the random variables $\left\{P_1^S(w_{[1:2]},f_{[1:2]}|z)\right\}$ are identically distributed for all $w_{[1:2]},f_{[1:2]}$ due to the symmetry in the random binning. Let $\bold{b}\triangleq\{b_1^{(j)},b_2^{(j)},j=1,2\}$ be a realization of the random binning $\setB$. We define the class $\setG$ of binning functions $\bold{b}$ as 
\begin{align}
\label{eq:appendix_B_27}
\setG\triangleq\left\{\bold{b}:{P}_1^S(w_{[1:2]},f_{[1:2]}|z)<\frac{1+\epsilon}{\tW_2\tF_2}P_{W_1F_1|Z_S}(w_1,f_1|z),
\text{ for all } S\in\setS \text{ and } z\in\setA_S\right\}.
\end{align}

Using the union bound, we have
\begin{align}
\label{eq:appendix_B_28}
\Prob_{\setB}(\setG^c)&=\Prob_{\setB}\left(P_1^S(w_{[1:2]},f_{[1:2]}|z)\geq\frac{1+\epsilon}{\tW_2\tF_2}P_{W_1F_1|Z_S}(w_1,f_1|z),\text{ for some } S\in\setS \text{ or } z\in\setA_S\right)\\
\label{eq:appendix_B_29}
&\leq \sum_{S\in\setS, z\in\setA_S} \Prob_{\setB}\left(P_1^S(w_{[1:2]},f_{[1:2]}|z)\geq\frac{1+\epsilon}{\tW_2\tF_2}P_{W_1F_1|Z_S}(w_1,f_1|z)\right)\\
\label{eq:appendix_B_30}
&\leq \sum_{S\in\setS}|A_S|\exp{\left(\frac{-\epsilon^2 (1-\delta) 2^{\gamma_{21}}}{3\tW_2\tF_2}\right)}\\
\label{eq:appendix_B_31}
&\leq |S||\setZ|\exp{\left(\frac{-\epsilon^2 (1-\delta) 2^{\gamma_{21}}}{3\tW_2\tF_2}\right)},
\end{align} 
where (\ref{eq:appendix_B_30}) follows from (\ref{eq:appendix_B_26}).

Take $\bold{b}$ such that $\bold{b}\in\setG$, and set $W_j=b_1^{(j)}(X_j)$ and $F_j=b_2^{(j)}(X_j)$ for $j=1,2$. For all $S\in\setS$, we have
\begin{align}
\label{eq:appendix_B_32}
&\nonumber\E_{p_{Z_S}}\left(\D\left(P_{W_{[1:2]}F_{[1:2]}|Z_S}||P_{W_1F_1|Z_S}p_{W_2}^U p_{F_2}^U\right)\right)\\
&=\E_{p_{Z_S}}\left(\sum_{w_{[1:2]},f_{[1:2]}}P_{W_{[1:2]}F_{[1:2]}|Z_S}(w_{[1:2]},f_{[1:2]}|Z_S)\log\frac{P_{W_{[1:2]}F_{[1:2]}|Z_S}(w_{[1:2]},f_{[1:2]}|Z_S)}{P_{W_1F_1|Z_S}(w_1,f_1|Z_S)p_{W_2}^U p_{F_2}^U}\right)\\
\label{eq:appendix_B_33}
&=\E_{p_{Z_S}}\left(\sum_{w_{[1:2]},f_{[1:2]}}\sum_{i=1}^2 P_i^S(w_{[1:2]},f_{[1:2]}|Z_S)\log\frac{\sum_{i=1}^2 P_i^S(w_{[1:2]},f_{[1:2]}|Z_S)}{\frac{P_{W_1F_1|Z_S}(w_1,f_1|Z_S)}{\tW_2\tF_2}\overset{2}{\underset{i=1}\sum}\;\;\underset{w_{[1:2]},f_{[1:2]}}\sum P_i^S(w_{[1:2]},f_{[1:2]}|Z_S)}\right)\\
\label{eq:appendix_B_34}
&\leq \E_{p_{Z_S}}\left(\underset{w_{[1:2]},f_{[1:2]}}\sum\sum_{i=1}^2 P_i^S(w_{[1:2]},f_{[1:2]}|Z_S) \log\frac{P_i^S(w_{[1:2]},f_{[1:2]}|Z_S)}{\frac{P_{W_1F_1|Z_S}(w_1,f_1|Z_S)}{\tW_2\tF_2}\underset{w_{[1:2]},f_{[1:2]}}\sum P_i^S(w_{[1:2]},f_{[1:2]}|Z_S)}\right)\\
\label{eq:appendix_B_35}
\nonumber&=\E_{p_{Z_S}}\left(\sum_{i=1}^2\sum_{w_{[1:2]},f_{[1:2]}} P_i^S(w_{[1:2]},f_{[1:2]}|Z_S)\log \frac{1}{\underset{w_{[1:2]},f_{[1:2]}}\sum P_i^S(w_{[1:2]}, f_{[1:2]}|Z_S)}\right)\\
\nonumber &\qquad +\E_{p_{Z_S}}\left(\sum_{w_{[1:2]},f_{[1:2]}} P_1^S(w_{[1:2]},f_{[1:2]}|Z_S)\log\frac{\tW_2\tF_2 P_1^S(w_{[1:2]},f_{[1:2]}|Z_S)}{P_{W_1 F_1|Z_S}(w_1,f_1|Z_S)}\right)\\
&\qquad +\E_{p_{Z_S}}\left(\sum_{w_{[1:2]},f_{[1:2]}} P_2^S(w_{[1:2]},f_{[1:2]}|Z_S)\log\frac{\tW_2\tF_2 P_2^S(w_{[1:2]},f_{[1:2]}|Z_S)}{P_{W_1 F_1|Z_S}(w_1,f_1|Z_S)}\right),
\end{align}
where (\ref{eq:appendix_B_34}) follows from the log-sum inequality. 

Now, we upper bound each term in the right hand side of (\ref{eq:appendix_B_35}) for $\bold{b}\in\setG$. The second term in the right hand side of (\ref{eq:appendix_B_35}) is upper bounded as follows:
\begin{align}
\label{eq:appendix_B_36}
\nonumber &\E_{p_{Z_S}}\left(\sum_{w_{[1:2]},f_{[1:2]}} P_1^S(w_{[1:2]},f_{[1:2]}|Z_S)\log\frac{\tW_2\tF_2 P_1^S(w_{[1:2]},f_{[1:2]}|Z_S)}{P_{W_1 F_1|Z_S}(w_1,f_1|Z_S)}\right)\\
\nonumber &=\E_{p_{Z_S}}\left(\sum_{w_{[1:2]},f_{[1:2]}} P_1^S(w_{[1:2]},f_{[1:2]}|Z_S)\log\frac{\tW_2\tF_2 P_1^S(w_{[1:2]},f_{[1:2]}|Z_S)}{P_{W_1 F_1|Z_S}(w_1,f_1|Z_S)}\mathbbm{1}\left\{Z_S\notin\setA_S\right\} \right)\\
&\quad +\E_{p_{Z_S}}\left(\sum_{w_{[1:2]},f_{[1:2]}} P_1^S(w_{[1:2]},f_{[1:2]}|Z_S)\log\frac{\tW_2\tF_2 P_1^S(w_{[1:2]},f_{[1:2]}|Z_S)}{P_{W_1 F_1|Z_S}(w_1,f_1|Z_S)}\mathbbm{1}\left\{Z_S\in\setA_S\right\}\right)\\
\label{eq:appendix_B_37}
\nonumber &\leq\log(\tW_2\tF_2)\;\E_{p_{Z_S}}\left(\sum_{w_{[1:2]},f_{[1:2]}} P_1^S(w_{[1:2]},f_{[1:2]}|Z_S)\mathbbm{1}\left\{Z_S\notin\setA_S\right\}\right)\\
&\quad+\E_{p_{Z_S}}\left(\sum_{w_{[1:2]},f_{[1:2]}} P_1^S(w_{[1:2]},f_{[1:2]}|Z_S)\log\frac{\tW_2\tF_2 P_1^S(w_{[1:2]},f_{[1:2]}|Z_S)}{P_{W_1 F_1|Z_S}(w_1,f_1|Z_S)}\mathbbm{1}\left\{Z_S\in\setA_S\right\}\right)\\
\label{eq:appendix_B_38}
\nonumber&\leq \log(\tW_2\tF_2)\sum_{z\in\setZ}p_{Z_S}(z)\mathbbm{1}\{z\notin\setA_S\}\sum_{w_{[1:2]},f_{[1:2]}}P_1^S(w_{[1:2]},f_{[1:2]}|z)\\
&\qquad\qquad +\log(1+\epsilon)\;\E_{p_{Z_S}}\left(\sum_{w_{[1:2]},f_{[1:2]}} P_1^S(w_{[1:2]},f_{[1:2]}|Z_S)\right)\\
\label{eq:appendix_B_39}
&\leq \Prob_{p_{Z_S}}\left(Z_S\notin\setA_S\right)\log(\tW_2\tF_2)+\log(1+\epsilon)\\
\label{eq:appendix_B_40}
&\leq \delta\log(\tW_2\tF_2)+\epsilon,
\end{align}
where (\ref{eq:appendix_B_37}) follows because, for $i=1,2,$ 
\begin{align}
\label{eq:appendix_B_41}
P_{i}^S(w_{[1:2]},f_{[1:2]}|Z_S)&\leq P_{W_{[1:2]} F_{[1:2]}|Z_S}(w_{[1:2]},f_{[1:2]}|Z_S)\\
\label{eq:appendix_B_42}
&=P_{W_1 F_1|Z_S}(w_1,f_1|z)P_{W_2 F_2|W_1 F_1 Z_S}(w_2,f_2|w_1,f_1,z)\\
\label{eq:appendix_B_43}
&\leq P_{W_1 F_1|Z_S}(w_1,f_1|z),
\end{align}
and hence $\frac{P_{i}^S(w_{[1:2]},f_{[1:2]}|Z_S)}{P_{W_1F_1|Z_S}(w_1,f_1|z)}\leq 1$
for all $w_{[1:2]},f_{[1:2]}$ and $i=1,2$. Equation (\ref{eq:appendix_B_38}) follows because, from (\ref{eq:appendix_B_27}), we have for all $\bold{b}\in\setG$ and $z\in\setA_S$ that $\frac{\tW_2\tF_2 P_1^S(w_{[1:2]},f_{[1:2]}|Z_S)}{P_{W_1 F_1|Z_S}(w_1,f_1|Z_S)}<(1+\epsilon)$. 

Next, we upper bound the third term in the right hand side of (\ref{eq:appendix_B_35}). We have that 
\begin{align}
\label{eq:appendix_B_44}
\nonumber\E_{p_{Z_S}}&\left(\sum_{w_{[1:2]},f_{[1:2]}}{P}_2^S(w_{[1:2]},f_{[1:2]}|Z_S)\right)\\
\nonumber &=\sum_z p_{Z_S}(z)\sum_{x_{[1:2]}} p_{X_{[1:2]}|Z_S}(x_{[1:2]},z)\mathbbm{1}\left\{\left(x_{[1:2]},z\right)\notin\setD_1^S\right\}\\
&\qquad\qquad\times\sum_{w_{[1:2]},f_{[1:2]}}\mathbbm{1}\left\{\setB_{1}^{(j)}(x_j)=w_j,\setB_{2}^{(j)}(x_j)=f_j,j=1,2\right\}
\\
\label{eq:appendix_B_45}
&=\Prob_{p_{X_{[1:2]},Z_S}}\left(\left(X_{[1:2]},Z_S\right)\notin\setD_{1}^S\right)\\
\label{eq:appendix_B_46}
&\leq \delta^2,
\end{align}
where (\ref{eq:appendix_B_46}) follows from the assumption of the lemma in (\ref{eq:lemma2_4}). Using (\ref{eq:appendix_B_43}) and (\ref{eq:appendix_B_46}), we have 
\begin{align}
\label{eq:appendix_B_47}
\nonumber \E_{p_{Z_S}}&\left(\sum_{w_{[1:2]},f_{[1:2]}} P_2^S(w_{[1:2]},f_{[1:2]}|Z_S)\log\frac{\tW_2\tF_2 P_2^S(w_{[1:2]},f_{[1:2]}|Z_S)}{P_{W_1 F_1|Z_S}(w_1,f_1|Z_S)}\right)\\
&\leq \log(\tW_2\tF_2)\;\E_{p_{Z_S}}\left(\sum_{w_{[1:2]},f_{[1:2]}} P_2^S(w_{[1:2]},f_{[1:2]}|Z_S)\right)\\
\label{eq:appendix_B_48}
&\leq \delta^2\log(\tW_2\tF_2).
\end{align}

Since we have 
\begin{align}
\label{eq:appendix_B_49}
&\sum_{i=1}^2\sum_{w_{[1:2]},f_{[1:2]}} {P}_i^S\left(w_{[1:2]},f_{[1:2]}|Z_S\right)=1,\\
\label{eq:appendix_B_50}
&\sum_{w_{[1:2]},f_{[1:2]}} {P}_1^S\left(w_{[1:2]},f_{[1:2]}|Z_S\right)=\Prob_{p_{X_{[1:2]}|Z_S}}\left(\left(X_{[1:2]},Z_S\right)\in\setD_{1}^S\right),\\
\label{eq:appendix_B_51}
&\text{and } \sum_{w_{[1:2]},f_{[1:2]}} {P}_2^S\left(w_{[1:2]},f_{[1:2]}|Z_S\right)=1-\Prob_{p_{X_{[1:2]}|Z_S}}\left(\left(X_{[1:2]},Z_S\right)\in\setD_{1}^S\right),
\end{align}
the first term on the right hand side of (\ref{eq:appendix_B_35}) is upper bounded as follows: 
\begin{align}
\label{eq:appendix_B_52}
\nonumber \E_{p_{Z_S}}&\left(\sum_{i=1}^2\sum_{w_{[1:2]},f_{[1:2]}} P_i^S(w_{[1:2]},f_{[1:2]}|Z_S)\log \frac{1}{\underset{w_{[1:2]},f_{[1:2]}}\sum P_i^S(w_{[1:2]}, f_{[1:2]}|Z_S)}\right)\\
&=\E_{p_{Z_S}}\left(H_b\left(\Prob_{p_{X_{[1:2]}|Z_S}}\left(\left(X_{[1:2]},Z_S\right)\in\setD_{1}^S\right)\right)\right)\\
\label{eq:appendix_B_53}
&\leq H_b\left(\E_{p_{Z_S}}\left(\Prob_{p_{X_{[1:2]}|Z_S}}\left(\left(X_{[1:2]},Z_S\right)\in\setD_{1}^S\right)\right)\right)\\
\label{eq:appendix_B_54}
&=H_b\left(\Prob_{p_{X_{[1:2]},Z_S}}\left(\left(X_{[1:2]},Z_S\right)\in\setD_{1}^S\right)\right)\\
\label{eq:appendix_B_55}
&\leq H_b(1-\delta^2)=H_b(\delta^2),
\end{align}
where (\ref{eq:appendix_B_53}) follows from Jensen's inequality and the concavity of $H_b$, and (\ref{eq:appendix_B_55}) follows from (\ref{eq:lemma2_4}) and that $H_b(x)$ is monotonically decreasing in $x\in(\frac{1}{2},1)$. 

Using (\ref{eq:appendix_B_40}), (\ref{eq:appendix_B_48}), and (\ref{eq:appendix_B_55}), for any $\bold{b}\in\setG$ and for all $S\in\setS$, the left hand side of (\ref{eq:appendix_B_35}) is upper bounded as
\begin{align}
\label{eq:appendix_B_56}
\E_{p_{Z_S}}\left(\D\left(P_{W_{[1:2]}F_{[1:2]}|Z_S}||P_{W_1F_1|Z_S}p_{W_2}^U p_{F_2}^U\right)\right)\leq \epsilon+(\delta+\delta^2)\log (\tW_2\tF_2)+ H_b(\delta^2)\leq \tilde{\epsilon}.
\end{align} 
Thus, the second probability on the right hand side of (\ref{eq:appendix_B_4}) is upper bounded as
\begin{align}
\label{eq:appendix_B_57}
\nonumber \Prob_{\setB}&\left(\max_{S\in\setS}\E_{p_{Z_S}}\D\left(P_{W_{[1:2]}F_{[1:2]}|Z_S}||P_{W_1F_1|Z_S}p_{W_2}^Up_{F_2}^U\right)>\tilde{\epsilon}\right)\\
&=1-\Prob_{\setB}\left(\max_{S\in\setS}\E_{p_{Z_S}}\D\left(P_{W_{[1:2]}F_{[1:2]}|Z_S}||P_{W_1F_1|Z_S}p_{W_2}^Up_{F_2}^U\right)\leq \tilde{\epsilon}\right)\\
\label{eq:appendix_B_58}
&=1-\Prob_{\setB}\left(\E_{p_{Z_S}}\D\left(P_{W_{[1:2]}F_{[1:2]}|Z_S}||P_{W_1F_1|Z_S}p_{W_2}^Up_{F_2}^U\right)\leq \tilde{\epsilon} \text{ for all } S\in\setS\right)\\
\label{eq:appendix_B_59}
&\leq 1-\Prob_{\setB}(\setG)=\Prob_{\setB}(\setG^c)\\
\label{eq:appendix_B_60}
&\leq |S||\setZ|\exp{\left(\frac{-\epsilon^2 (1-\delta) 2^{\gamma_{21}}}{3\tW_2\tF_2}\right)},
\end{align} 
where (\ref{eq:appendix_B_60}) follows from (\ref{eq:appendix_B_31}).

Finally, by rewriting (\ref{eq:appendix_B_3}) with switching the roles of $(W_1,F_1)$ and $(W_2,F_2)$ and repeating the whole proof, we obtain the second term in the minimum in (\ref{eq:lemma2}), which completes the proof for Lemma \ref{lemma2}.

\bibliographystyle{IEEEtran}
\bibliography{MyLib}

\end{document}